\documentclass[aps,superscriptaddress,preprint]{revtex4-1}
\usepackage{amsmath}
\usepackage{amssymb}
\usepackage{array}
\usepackage{adjustbox}
\usepackage{color}
\usepackage{graphicx}
\usepackage{latexsym}
\usepackage{natbib}
\usepackage[caption=false]{subfig}
\usepackage{multirow}
\usepackage{mathtools}
\usepackage{textcomp}
\usepackage{rotating}
\usepackage{tabularx}

\usepackage{xurl}
\usepackage{longtable}
\usepackage[breaklinks=true]{hyperref}

\newcommand{\SI}{ Supplementary Information }

\usepackage[final]{feynmp}
\begin{document}

\title{Why polls fail to predict elections}

\author{Zhenkun Zhou}
\affiliation{College of Statistics, Capital University of\\ 
    Economics and Business, Beijing 100070, China}
\author{Matteo Serafino}
	\affiliation{IMT School for Advanced Studies, 55100 Lucca, Italy}
\author{Luciano Cohan}
\affiliation{Seido, Buenos Aires, Argentina}
\author{Guido Caldarelli}
\affiliation{Department of Molecular Sciences and Nanosystems,
Ca' Foscari University of Venice, 30172 Venice, Italy}
\affiliation{European Centre for Living Technology, 30124 Venice, Italy}
\affiliation{Institute for Complex Systems, Consiglio Nazionale
delle Ricerche, UoS Sapienza, 00185 Rome, Italy}
\affiliation{London Institute for Mathematical Sciences,
W1K2XF London, United Kingdom}
\author{Hern\'an A. Makse}
	\email{hmakse@ccny.cuny.edu}
  \affiliation{Levich Institute and Physics Department,\\
  City College of New York, New York, NY 10031, USA}

  \begin{abstract}
    
In the past decade we have witnessed the failure of traditional polls
in predicting presidential election outcomes across the world. To
understand the reasons behind these failures we analyze the raw data
of a trusted pollster which failed to predict, along with the rest of
the pollsters, the surprising 2019 presidential election in Argentina
which has led to a major market collapse in that country.  Analysis of
the raw and re-weighted data from longitudinal surveys performed
before and after the elections reveals clear biases (beyond well-known
low-response rates) related to mis-representation of the population
and, most importantly, to social-desirability biases, i.e., the
tendency of respondents to hide their intention to vote for
controversial candidates. We then propose a longitudinal opinion
tracking method based on big-data analytics from social media, machine
learning, and network theory that overcomes the limits of
traditional polls. The model achieves accurate results in the 2019
Argentina elections predicting the overwhelming victory of the
candidate Alberto Fern\'andez over the president Mauricio Macri; a
result that none of the traditional pollsters in the country was able
to predict. Beyond predicting political elections, the framework we
propose is more general and can be used to discover trends in
society; for instance, what people think about economics, education or
climate change.

  \end{abstract}


\maketitle

Traditional polling methods \cite{surveys} using random digit dial
phone interviews, opt-in samples of online surveys, and interactive
voice response are failing to predict election outcomes across the
world
\cite{kennedy2018evaluation,durand2020quebec,bekele2018faulty}. The
failure of traditional surveys has also been widely discussed in the
press \cite{2019Economist} and on the specialized literature
\cite{kennedy2018evaluation}. For instance, the victory of Donald
Trump in the US 2016 presidential election came as a shock to many, as
none of the pollsters and political journalists and pundits, including
those in Trump's campaign, could predict this victory
\cite{jacobs,kennedy2018evaluation}.

The reasons for the failure of pollsters to predict elections are
believed to be many \cite{kennedy2018evaluation,jacobs}. First reason
is that the percentage of response to traditionally conducted surveys
has decreased and it is becoming increasingly difficult to get
people's opinion
\cite{kennedy2018evaluation,kennedy2019response}. Response rates in
telephone polls with live interviewers continue to decline, as it has
reached 6\% lower limit recently \cite{kennedy2019response}. Response
rates could be even lower for other methodologies, like internet
polling or interactive voice response.  Compounded with declining
response rates is the concomitant problem of mis-representation of the survey samples. That is, the sample surveyed by
pollsters does not represent the demographic distributions of the
general population. This problem is ameliorated by reweighting the
surveys sample according to the general demographics of the population
in a process called sample-balancing or raking \cite{w1,w2}. However,
in countries where the vote is not obligatory, re-weighting a sample
to the general population demographics (obtained from Census Bureau
\cite{kennedy2018evaluation,surveys}) fails since the general
population demographics does not match necessarily the demographics of
the voter turnout: it is impossible to predict which
demographic groups will turn out at the voting station. Thus, if an
underrepresented group in the polls, {\it 'the hidden vote'}, decides
to vote on election date, the re-weighting fails leading to highly
inaccurate results. The issue is believed to be one of the major
reasons for the generalized failure of pollsters to predict the
triumph of Trump in the 2016 US presidential election, where groups
generally defined as 'white voters without college degree' mostly
voted for Trump but were undersampled by all pollsters. Even with this
historical information at hand, which supposedly allowed pollsters to
resample their surveys more carefully, pollsters again under-predicted
the support for Trump in the subsequent 2020 presidential election in
some states or under-predicted the voter turnout supporting Biden in
newly created battleground states like Georgia, USA \cite{nyt1}. The
inability to accurately predict the voter turnout to deal with
sampling mis-representation might render the pollsters obsolete.

While there is increasing evidence
\cite{kennedy2018evaluation,kennedy2019response} that the nonresponse
and mis-representation bias might be the reason that polls are not
producing accurately matched election results, these may not be the
only problem of traditional methods of polling. Traditional surveys in
heavily polarized campaigns are affected by social-desirability biases
(also called Bradley effect) \cite{Payne,Krumpal}, i.e. the tendency
of subjects to give socially desirable responses instead of choosing
responses that are reflective of their true feelings. For instance,
the tendency of survey respondents not to tell the truth of intention
of support for controversial candidates which could open themself to
social ostracism. Respondents may feel under pressure to provide
'politically correct' answers producing highly-biased results towards
the publically accepted candidate by the media, in detriment of the
controversial one.  This mechanism is also believed to have been at
place in the massive failure of pollsters to predict 2016 US election
as Trump voters generally can go undetected or even lie to
traditional pollsters.  We will show below, that it was also the major
reason for the pollsters failure to predict the 2019 Argentina
election, which also involved a controversial candidate (Cristina
Fern\'andez) who was heavily under-predicted in traditional
polls. Furthermore, polls are not able to detect sudden change of
opinion due to some particular events or circumstances, since the
process of opinions collection is time consuming. All these
peculiarities together makes impossible for traditional polls to
correctly predict the results of elections.

Monitoring social networks \cite{niaki,KBS01} represents an alternative for
capturing people's opinions since it overcomes the low-response rate
problem and it is less susceptible to social-desirability biases
\cite{Payne,Krumpal}. Indeed, social media users continuously express
their political preferences in online discussions without being
exposed to direct questions. One of the most studied social networks
is the microblogging platform Twitter
\cite{Jaidka,Jungherr,bovet,fake}. Twitter's based work generally
consist of three main steps: data collection, data processing and data
analysis. The collection of the tweets is often based on the public
API of Twitter. It is a common practice to collect tweets by filtering
according to specific queries, as for example the name of the
candidates in the case of elections \cite{bovet}. Data processing
includes all those techniques which aim to guarantee the credibility
of the Twitter dataset. This is, for example, bots detection and spam
removal \cite{fake}. Data analysis, the core of all these studies, can
be simplified in four main approaches: volume analysis, sentiment
analysis, network analysis and artificial intelligence (AI)
\cite{niaki}.

Scholars used the number of mentions for a party of a candidate in
order to forecast the result of the 2009 German parliament election
\cite{Tumasjan}. While their technique has attracted many criticisms
\cite{jungherr2012pirate}, their work was of inspiration to many other
researchers. Gaurav {\it et al.} \cite{Gaurav} proposed a model, based
on the number of times the name of a candidate is mentioned in tweets
prior to elections, to predict the winner of three presidential
elections held in Latin America (Venezuela, Paraguay, Ecuador) from
February to April, 2013. Based on volumetric analyses are also the
works of Lui {\it et al.} \cite{Lui} and Bermingham
\cite{Bermingham}. Ceron {\it et al.} \cite{ceron2014every} performed
a sentiment analysis study on the tweets to check the popularity of
political candidates in the Italian parliamentary election of 2011 and
in the French presidential election of 2012.  Caldarelli {\it et al.}
\cite{Caldarelliplosone} used the derivative of the volume to forecast
the results of Italian elections. Singh {\it et al.}
\cite{singh2017forecasting} employed sentiment analysis to predict
victory of Trump in the election of 2016. The same author proposed a
method \cite{singh2020can} based on sentiment analyses and machine
learning on historical data to predict the number of seats that
contesting parties were likely to win in the Punjab election of
2017. Other works \cite{Newman,Cuzzocrea} used social networks
analyses in order to identify the position of a party in the online
community by measuring its centrality. The most supported parties are
in general those with an higher centrality. Bovet {\it et al.}
\cite{bovet,fake} used a machine learning model based on in-house
training set produced by hashtags from Twitter supporters to reproduce
the polling trends leading to the 2016 US presidential election.

Despite the large amount of literature, the debate about whether
Twitter or other social media outlets can be used to infer political
opinions is still open. Online social networks are continuously filled
by false, erroneous data through trolls, bots and misinformation
campaigns to a level that distinguish between what is genuine and what
is not is in general difficult \cite{fake,KBS02}. By virtue of this, the
great challenge of algorithms and AI is to discover and interpret real
data from `junk data' that could lead to accurate predictions of
electoral or opinion trends. A crucial limitation of social media
based methods is also the mis-representation bias. While social media
solves the low response rates by 'surveying' millions of users with
non-intrusive methods, these large number of respondents might not
represent, again, the demographics of the voting population. Thus, the
opinions of Twitter users may not be representative of the entire
population \cite{bode2016politics} and re-sampling methods need to be
used, importing along them the same problems that plagued the
traditional polls.

In this work we first investigate why the traditional polls fail to
predict elections. We focus on the results of the recent primary
presidential election in Argentina on August 2019 and the subsequent
presidential election on October 2019, which represents a classic
example of a massive failure of the trusted pollsters in predicting a
polarized election electorate, which in this case, led also to massive
markets collapses in the country, since investors largely bet on the
pollster predictions.

This study is possible thanks to the exclusive access to the raw data
of longitudinal surveys conducted by one of the most reliable
pollsters in Argentina, Elypsis \cite{erraron}.  The analyzed data
include the original responses of subjects before performing the
re-weighting for sampling bias and the subsequent results obtained
after re-weighting. More importantly, the data includes a longitudinal
study on the same 1,900 respondents before and after the election
which allows to precisely study the social-desirability bias when the
same voter change the response after the result of the election is
known.  This represents an unique opportunity to discover why the
traditional polls have failed as pollsters do not normally share their
raw data before re-weighting or sample-balancing \cite{w1,w2} and few
results have been performed on the same respondents before and after
an election.  The raw data of this pollster firm have been obtained by
exclusive arrangement with the pollster responsible for conducting the
polls of Elypsis (co-author Luciano Cohan) who has later founded his
own company (Seido).

We find that a poor demographic representation combined with the
inconsistency of opinion' respondents before and after the elections
are the main reasons of the polls failure. We find a large
mis-representation of the sample in the surveys as compared with the
voting population, which in Argentina is the general population since
voting is obligatory and voter turnout is quite high at +80\%.  Even
after re-weighting, this large sample bias produces highly inaccurate
results since important segments of society are highly
underrepresented in the polls. Beyond this sampling problem, the main
problem we find is a clear tendency for the respondents to not tell
the truth about their preference for a candidate (Fernandez) who was
controversial and highly underdog in all polls and the media. This
social-desirability bias was the main culprit for the failure of the
polls.

To overcame these problems, we propose an AI model to predict
electorate trend using opinions extracted from social media like
Twitter.  By using machine learning first developed in \cite{bovet} we
uncover political and electoral trends without directly asking people
what they think, but trying to predict and interpret the enormous
amount of data they produce in online social media
\cite{bovet,fake,book,jasny}. Thus, these big-data analysis overcomes
the low response rate problem. By re-weighting the Twitter populations
to the Census data, we match the distribution of the population'
statistics and the statistics of the real population (given by the
Census Bureau) \cite{bode2016politics} thus minimizing the sampling
bias of Twitter. Since social media users freely express their
opinions in social media and our methods are not interventionist, the
data are, in principle, free of social-desirability bias. The real
time data processing which underlies our AI algorithm allows us to
detect sudden change of opinions, and therefore different loyalty
classes towards each candidate.

We will show that a cumulative longitudinal analysis tracking users
over time performed on the loyalties classes to the candidates
considerably improves previous results of~\citep{bovet}, based on
instantaneous predictions. Instantaneous predictions, as well as
pollsters predictions, are subject to high fluctuations which
undermine the reliability of the prediction itself. Instead, here we
show that taking into account the cumulative opinions of users over a
long period of time produces a reliable predictor of people's
opinion. These improvements allow us to obtain an accurate prediction
on a difficult election, which dodged all pollsters in
Argentina. Thus, we validate the algorithm on the primary and general
election in Argentina.  Our results in this particular case show that
AI can capture the public opinion more precisely and more efficiently
than traditional polls.

\section{Why pollsters are failing to predict elections?}

The events leading up to the recent primary election in Argentina are
a telling example of the failure of the polling industry
\cite{wiki-election,wiki-polls,erraron}. On the primary election day
on August 11, 2019 (called PASO in Spanish: Primarias, Abiertas,
Simult\'aneas y Obligatorias; in English: Open, Simultaneous, and
Obligatory Primaries), none of the pollsters in the country predicted
the wide 16\% margin of presidential candidate Alberto Fern\'andez
(AF) over the president Mauricio Macri (MM) [We clarify that primaries
  in Argentina are obligatory, happening for all political parties at
  the same time, and the two main parties presented only one
  candidate each, thus transforming the primaries into a de-facto
  presidential contest.]

Figure \ref{fig:comparison} shows the comparison between the official
results (in red), our prediction (in blue, Model 3
explained below) and the polling average, computed as the average of
the top five most trusted pollsters in Argentina
\cite{wiki-polls,erraron}, i.e. Real Time Data, Management \& Fit,
Opinaia, Giacobbe and Elypsis (in green).  Macri was clearly defeated
by Fernandez by +16\%, a result captured by our predictions.  While
the average pollster predicted Fernandez with a slight advantage in
the primary, the estimated percentage of each candidates were, in
general, really close reaching in some occasion a difference of just
one percentage point \cite{clarin}.  Elypsis in particular predicted
that Macri would win for one percentage point \cite{manipulation}.
This virtual tie predicted by the pollsters was largely considered to
be a win for the incumbent candidate Macri since he was supposed to
gain all the votes left by the third party options in the subsequent
presidential election and eventually win the election in a runoff.

It is worth to stress at this point that Macri (right-leaning
candidate) made of the internationalization of the market one of the
main points of his campaign (favoring foreign investors and
pro-business) while Fern\'andez (left-leaning candidate) was instead
supporting a national market. As a result of the predictions supported
by all Argentinian pollsters giving Macri as the winner, the bond
market rose excessively in the days preceding the primary
election. The subsequent defeat of Macri by 16 percentage points at
the primaries leads to a historic collapse of the MERVAL index by
40\%, the bond market collapsed and some banks lost 1 billion dollars
in the bet overnight \cite{wsj,ft}.
 
The failure of traditional polls is not associated with the
impossibility of giving the exact right percentage for the candidates,
but rather with the impossibility of predicting the enormous gap
between them. The Argentina primaries elections are not the only
example of pollsters' failure. Unpredictable results seems to be
associated whenever one of the candidates is a controversial figure in
the political scenario. In Argentina, the eventual vice-presidential
candidate accompanying Alberto Fern\'andez was previous Argentinian
president Cristina Fern\'andez de Kirchner (CFK), who had faced
corruption allegations and judicial processes and many Argentinians
and the traditional media viewed as a controversial and divisive
figure in Argentina politics \cite{post} and who is usually vilified
by the traditional media. Notorious examples are Trump in the American
presidential election of 2016 or Bolsonaro in the Brazilian general
election of 2018. In the case of the Argentina primaries, the pollster
failure led to economic disruption of the country at a
national/international level \cite{wsj}.

Below we analyze the raw data of one of the most reliable polls,
Elypsis (trusted specially by the president Macri and international
investors \cite{wsj,ft,erraron,clarin}) which as all the pollsters
failed to predict the large gap between the two candidates for the
primaries elections.

\begin{figure}[ht]
  \centering
  \subfloat[]{\includegraphics[width=.6\textwidth]{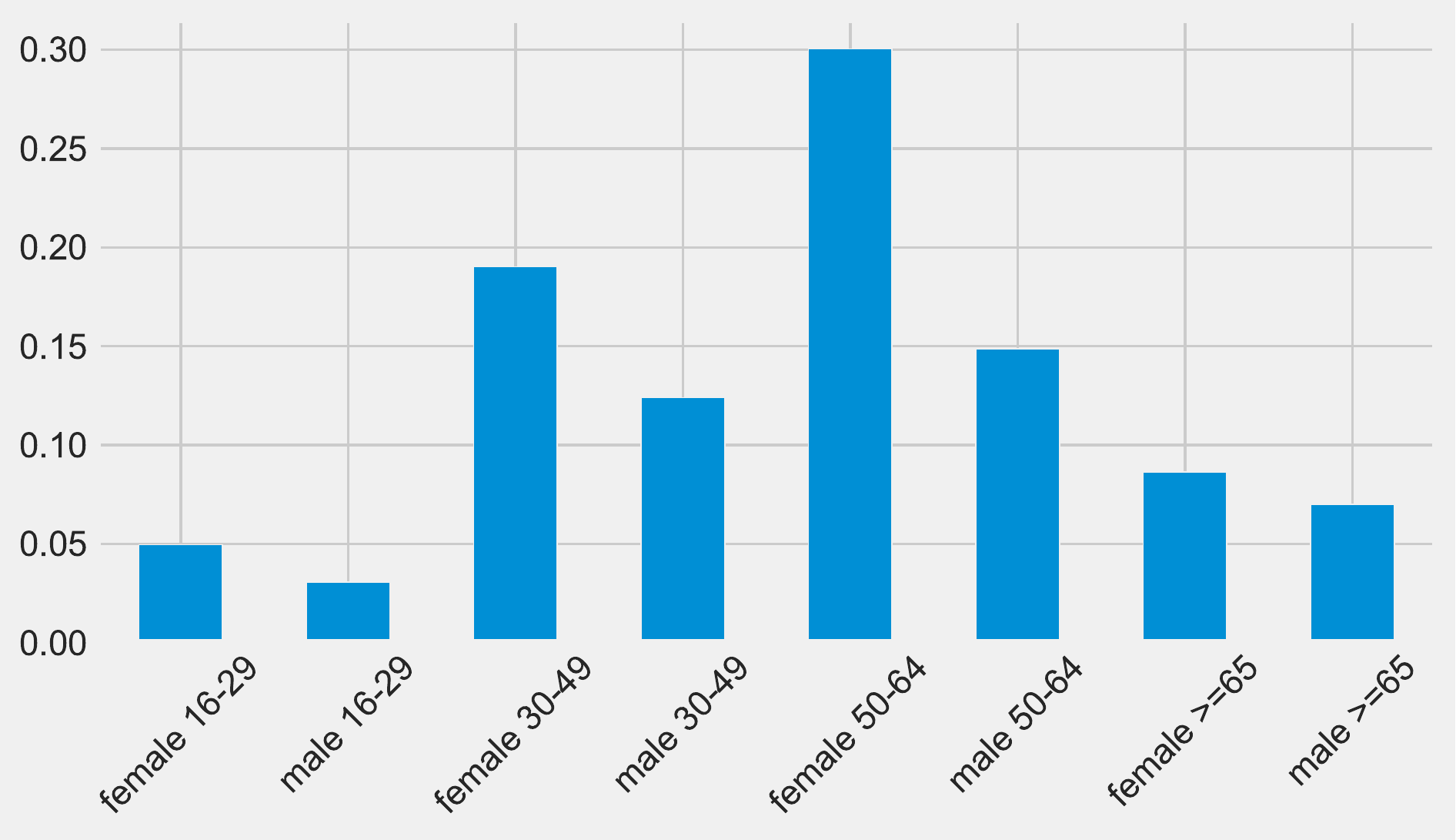}} \hfil
  \subfloat[]{\includegraphics[width=.6\textwidth]{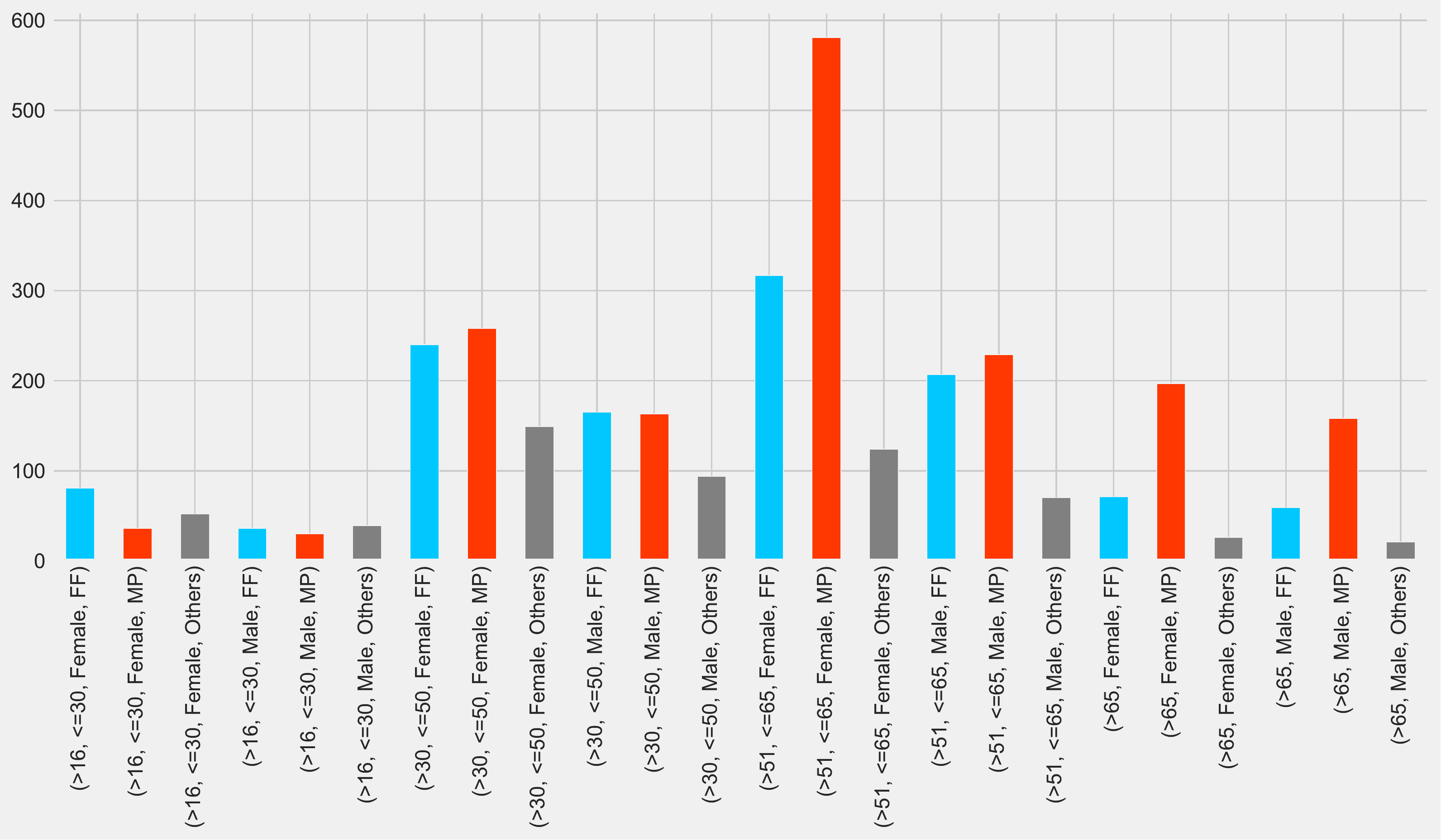}} \hfil
  \subfloat[]{\includegraphics[width=.6\textwidth]{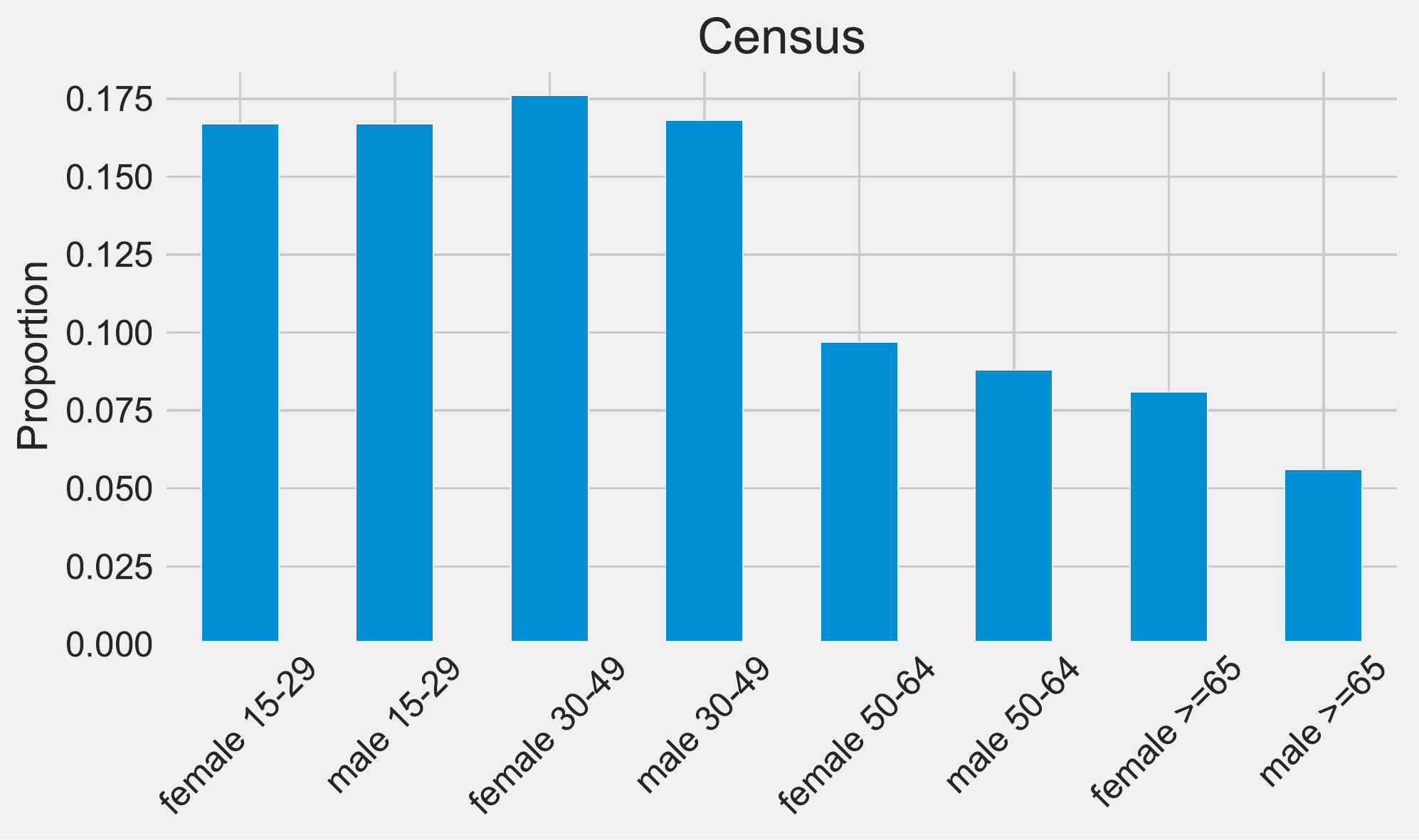}} \hfil
  \caption{
  (a) Elypsis demographics before PASO.
  (b) Elypsis polling results before PASO separated by age and gender (Macri=red, Fern\'andez= blue, Others=grey). Results are highly biases to older than 50 as compared with Census distribution.
  (c) Argentina Census Bureau 2010 demographic distribution by age and gender.}
 \label{fig:elypsis-population}
\end{figure}

Figure ~\ref{fig:elypsis-population}a shows the age-gender
distribution of the respondents to the survey conducted by Elypsis
immediately before the PASO elections. Elypsis employs a combination
of IVR of landline numbers complemented with online opt-in samples
from Facebook. The vast majority of these online panels in Argentina,
as well as in US, are made up of volunteers who were recruited online
and who received some form of compensation for completing surveys,
such as small amounts of money or frequent flyer
miles. Fig. \ref{fig:elypsis-population}b shows the number of
respondents for Fern\'andez (light blue), Macri (red) and Third Party
(grey) grouped by age and gender. The Elypsis sample is peaked around
the 50 years old group. This is strikingly different from the national
population statistics obtained from the Argentinian Census Bureau
shown in Fig. \ref{fig:elypsis-population}e.

Elypsis data does not have significant coverage among people younger
than 30, even though it has been conducted in Facebook. It shows a
heavier tail on the right for older groups, while the national
population (Census Bureau) has a less pronounced peak around the group
of 30 years old and an heavier tail on the left for younger
groups. The largest sampled group surveyed by Elypsis are females
between 51 and 65 years old who are overwhelming in favor of Macri. In
fact, in all groups above 30 years old, Macri is the clear favorite in
the Elypsis poll. On the contrary, Twitter represents better the
younger generations. It is important to consider again that the vote
is obligatory in Argentina and it is permitted above 16 years, and the
turnout of the youngest is quite substantial, thus, any pollster that
does not capture their preferences is, in practice, doomed to fail.
  
To deal with the mis-representation problem, pollsters adjust their
raw results to population benchmarks distributions given by the Census
Bureaus \cite{kennedy2018evaluation,surveys} by weighting the raw data
(sample-balancing or raking \cite{w1,w2}). The poll sample is weighted
so it matches the population on a set of relevant demographic or
political variables, for instance, age, gender, location and other
socio-economic variables, like education level or income. Studies of
the effectiveness of various weighting schemes suggest they reduce
some (30 to 60\%) of the error introduced by the biased sample, see
\cite{surveys}. However, when the raw data distribution is drastically
under/over sampled as the Elypsis case, a small error in the most
representative groups would propagate to produce inaccurate result.

\begin{figure}[!ht]
  \centering
  \subfloat[]{\includegraphics[width=.6\textwidth]{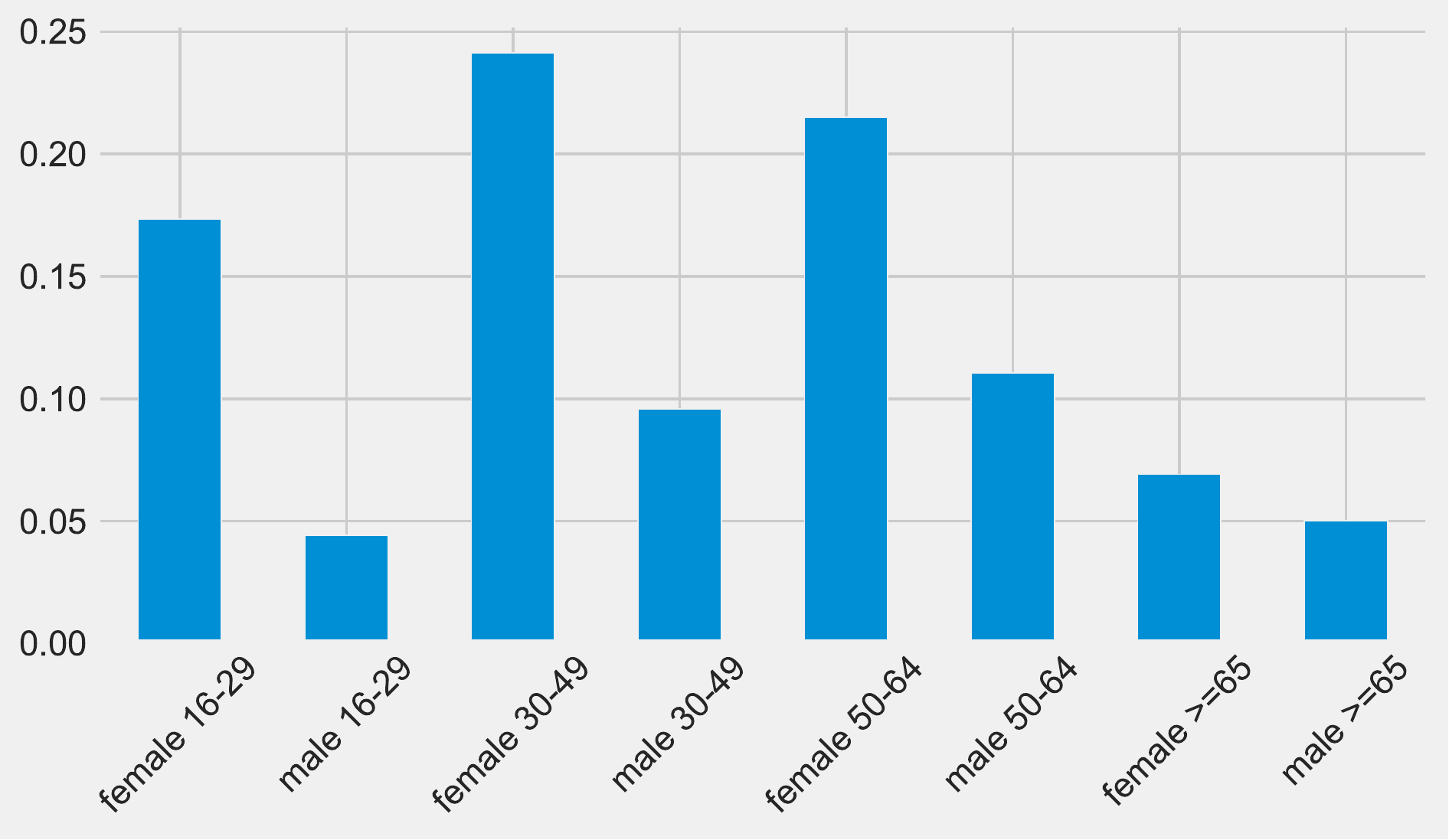}} \hfil
  \subfloat[]{\includegraphics[width=.6\textwidth]{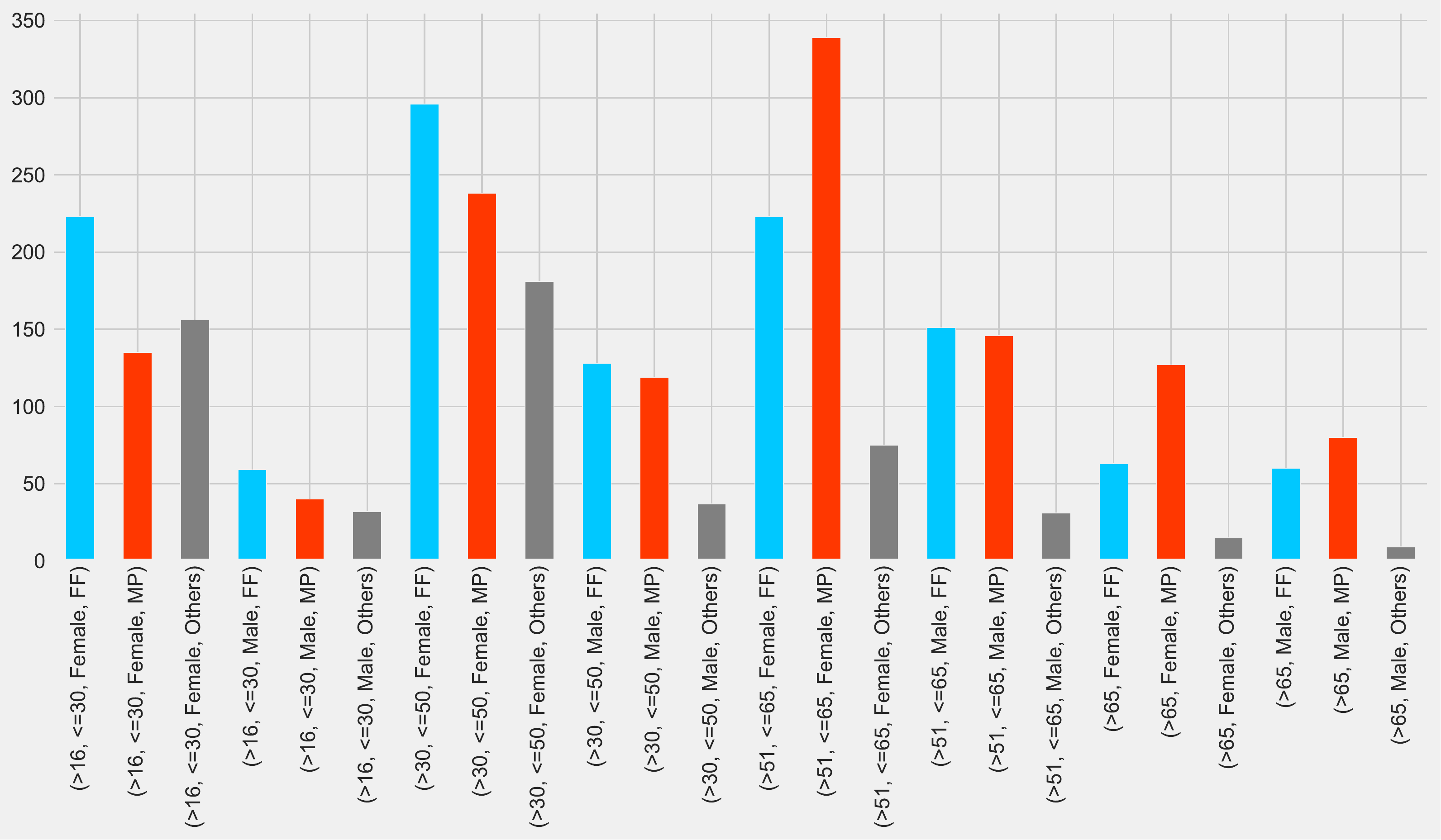}} \hfil
  \caption{
  (a) Elypsis demographics after PASO.
  (b) Elypsis polling results after PASO separated by age and gender  (Macri=red, Fern\'andez= blue, Others=grey).}
 \label{fig:elypsis-population_after}
\end{figure}

As discussed above, the mis-representation is not the only problem
which traditional pollsters methods face. Next, we analyze the
longitudinal data taken on the same 1,900 respondents by Elypsis
before and after the elections to investigate the social-desirability
bias. We start with Fig. \ref{fig:elypsis-population_after}a showing the
Elypsis respondent distributions after PASO (notice that these respondents 
from the previous one, and this is the reason why the age distribution 
change respect to the previous figure). By comparing
Fig. \ref{fig:elypsis-population}a before PASO with
Fig. \ref{fig:elypsis-population_after}a after PASO we first notice a change
in the voters distributions. Younger groups are better represented
after the election when compared to
Fig. \ref{fig:elypsis-population}a, although the data are still highly
biased towards older generations. This implies that younger groups
were, at least, more prone to answer the polls after the election than
before.

Surprisingly, the female group with ages between 30 and 50 years
voted for Fern\'andez as indicated after the PASO polls, while before
the PASO they responded mainly in favor of Macri. The male group
of the same age shows a similar behavior, even if less pronounced. Let
us notice that, according to Fig. \ref{fig:elypsis-population}b, the
groups of females/males between 30 and 50 years old are the most
represented in the Census data and therefore may have an higher impact
on the final result. These results can only be explained by admitting
that voters did not say the true.

This is further corroborated by this unique longitudinal panel, as
seen in Table \ref{table-longitudinal}, revealing that people lied and
hid their true voting intentions to the pollsters before the
elections.
More specifically, when comparing ``Who are you going to vote in the
PASO'' with ``Who did you vote in the PASO'' - using the same sampling
and postratification methodology than in the Pre-PASO survey - it is
found that about 18\% of the people did not disclose their true vote,
and the hidden vote was not unbiased.

\begin{itemize}
  
\item  91\% of those who said ``I will vote for Fernandez'' did so, but only
83\% in the case of Macri, who lost 6\% to AF.

\item ``Secondary candidates'' voters were much more volatile, Only 56\%
of those who said that they were going to vote for (third candidate)
Lavagna disclosed their true vote, and 54\%, 53\% and 59\% in the case
of other candidates Del Ca\~no, Espert and Gomez Centurion
respectively.

\item  Alberto Fern\'andez got almost 19\% of the votes of those who chose a
secondary candidate in the Pre Paso Poll, and Mauricio Macri only 9\%.

\item  Alberto Fern\'andez received 46\% of the votes of those who answered
"Blank, Null or Unknown" before the PASO.

\end{itemize}

But, who hid - or not disclosed - their real vote? We find no
significant difference between men and women or between education
levels but we see a clear pattern in age demographics. 33\% of those
between 16 and 30 years changed their vote vs. their Pre PASO answer
and only 13\%, 10\% and 14\% on those between 31 and 50, 51 and 65 and
more than 65, see Table \ref{table2}.

What did those who did not disclose their vote think about the
candidates? Where they "closeted Kirchnerists" (party of AF and CFK)
or did they bridge the gap between Macri-Fern\'andez?

"Regular" images of Cristina Fern\'andez, Macri and
Alberto Fern\'andez  were much lower among those that did reveal their
vote than among those who did not, see Table \ref{table3}. Those who
hid their votes look more nonpolarized, with a "Regular" image - No
positive nor negative - of 21\% on average, vs 6\%/10\% of those who
revealed the vote.  CFK's negative image is higher than MM (48\% vs
38\%) in "non-revealers" and the opposite hold in the revealers (43\%
vs 50\%).  35\% of the "non-revealers" did not have (or hid) their
opinion of Alberto Fernandez vs. 8\% in the revealers.

This combined information shed light on PASO results and Polls
consensus miss. In the PASO, AF was able to catch votes from all the
candidates, and seduce voters from within the gap, "moderate" voters
who had a negative image of CFK and MM. He succeeded in standing
himself as the "third candidate" bridging the gap, something that was
not being fully captured by the polls, or that was decided at the last
minute. This feature is most striking in young people, who may both
have more "volatile" opinions and less prone to reveal them on
traditional polls.

This hidden-vote factor can explain by itself as much as 10\%
difference between "ex-ante" forecast and real results. Thus, standard
polls methods failure may not have been related only to a bias in
the sampling but, in the extraction of "True" information from
surveyed people.

Understanding why people lie is not the topic of this work even if,
according to the literature the reasons could be many and related to
desirability-bias. On one hand, participants may typically rush
through the surveys to obtain their rewards and don't respond
thoughtfully \cite{kennedy2018evaluation}. On the other hand,
social-desirability bias \cite{Payne,Krumpal}, i.e. the tendency of
survey respondents to answer questions in a manner that will be viewed
favorably by others \cite{Krumpal,kennedy2018evaluation} is another
reason for people to hide their preference for controversial
candidates like CFK, which leads to biased results.

In view of how the above issues of low response rate,
mis-representation and the social desirability bias/lies (which in the
case of Elypsis biased more the younger representative) undermined the
predictions on the Argentinian primary elections, we next search for
suitable replacement using sampling methods for the modern era of
big-data science. In this scenario, a good candidate to substitute
traditional polls is social media (Twitter in our study) which solves
in one shot both the law response rate (million of people express
their political preferences in the microblogging platform) and the
social desirability biases. This is because social media users do not
answer to any question, but freely express their ideas in a social
medium platform.  However, one may argue that Twitter is generally
bias towards young people thus providing a biased sample. Thus, proper
re-weighting of the data is needed, although the effects of
re-weighting are expected to be less pronounced than in the polls of
Elypsis. Below we introduce an AI model that builds up on previous
work in \cite{bovet} combining machine learning, network theory and
big-data analytic techniques, that is able to overcome the problems
presented so far and that correctly predicted the outcome of the 2019
Argentina primary and general elections.

\section{Methodology}

\begin{figure}[!ht]
  \includegraphics[width=.9\textwidth]{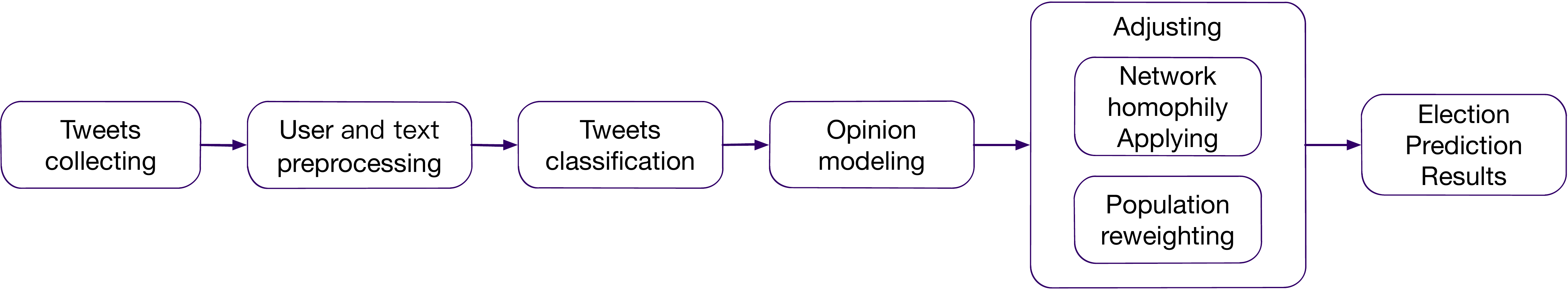}
  \caption {The flow of election prediction algorithm.}
 \label{fig:flow}
\end{figure}

The algorithm we propose improves upon previous work from \cite{bovet}
and consists of four phases (see Fig.~\ref{fig:flow}): data
collection, text and user processing, tweets classification with
machine learning and opinion modeling. While the first two phases are
of standard practice in the literature, tweets classification by means
of ML models only recently took place \cite{bovet,fake}, given the
impossibility to classify by hand millions and millions of
data. Opinion modeling, the core of our election prediction model, is
an attempt to instantly capture people's opinion through time by means
of a social network. To improve upon \cite{bovet}, we consider the
cumulative opinion of people and define five prediction models based
on different assumptions on the loyalty classes of users to
candidates, homophily measures and re-weighting scenarios of the raw
data. Below we explain each phase, highlighting the steps that make
our full-fledge AI predictor a good candidate substitute for the
traditional pollster methods.

\begin{figure}
  \centering
  \subfloat[]{\includegraphics[width=.5\textwidth]{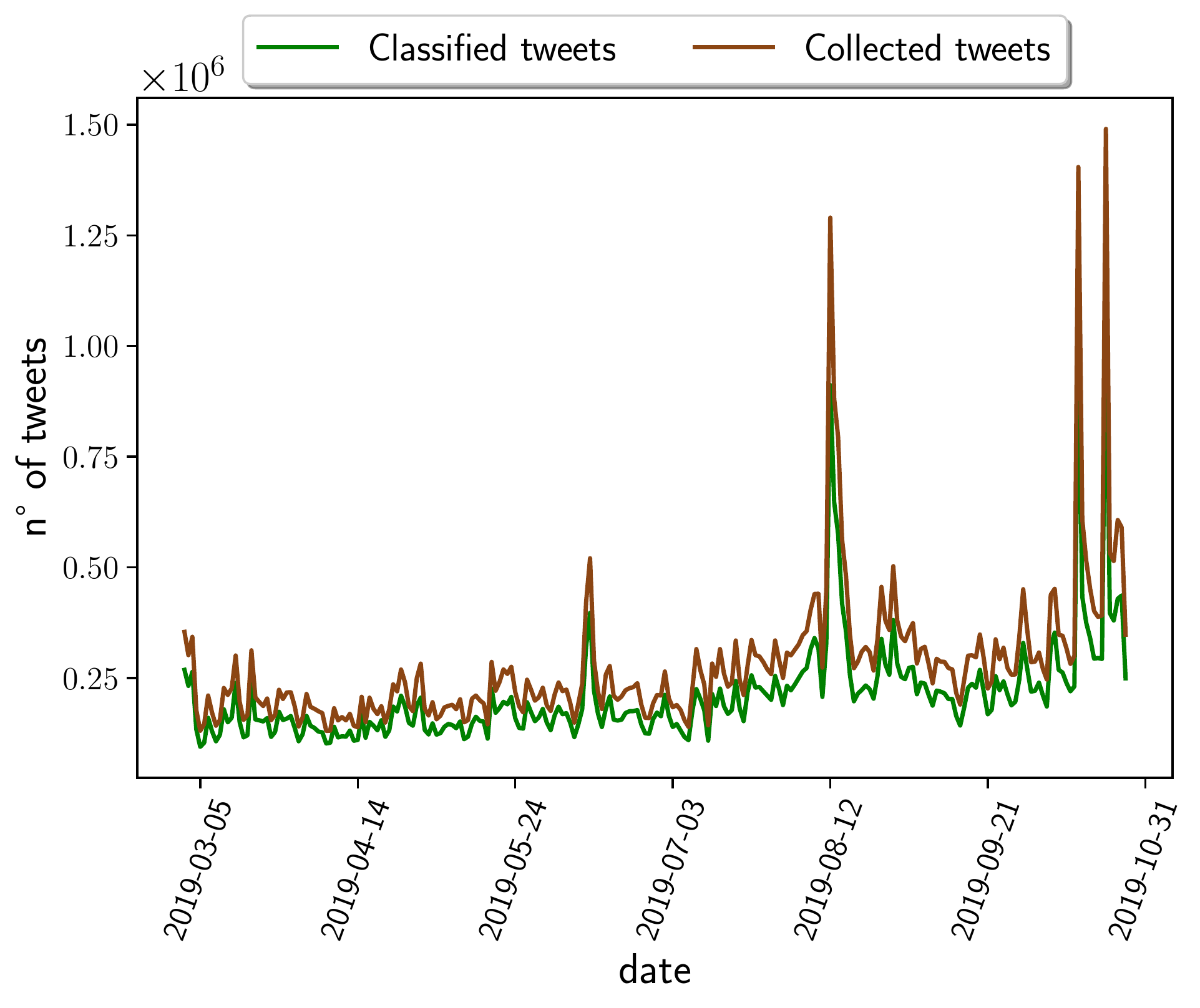}} \hfil
  \subfloat[]{\includegraphics[width=.5\textwidth]{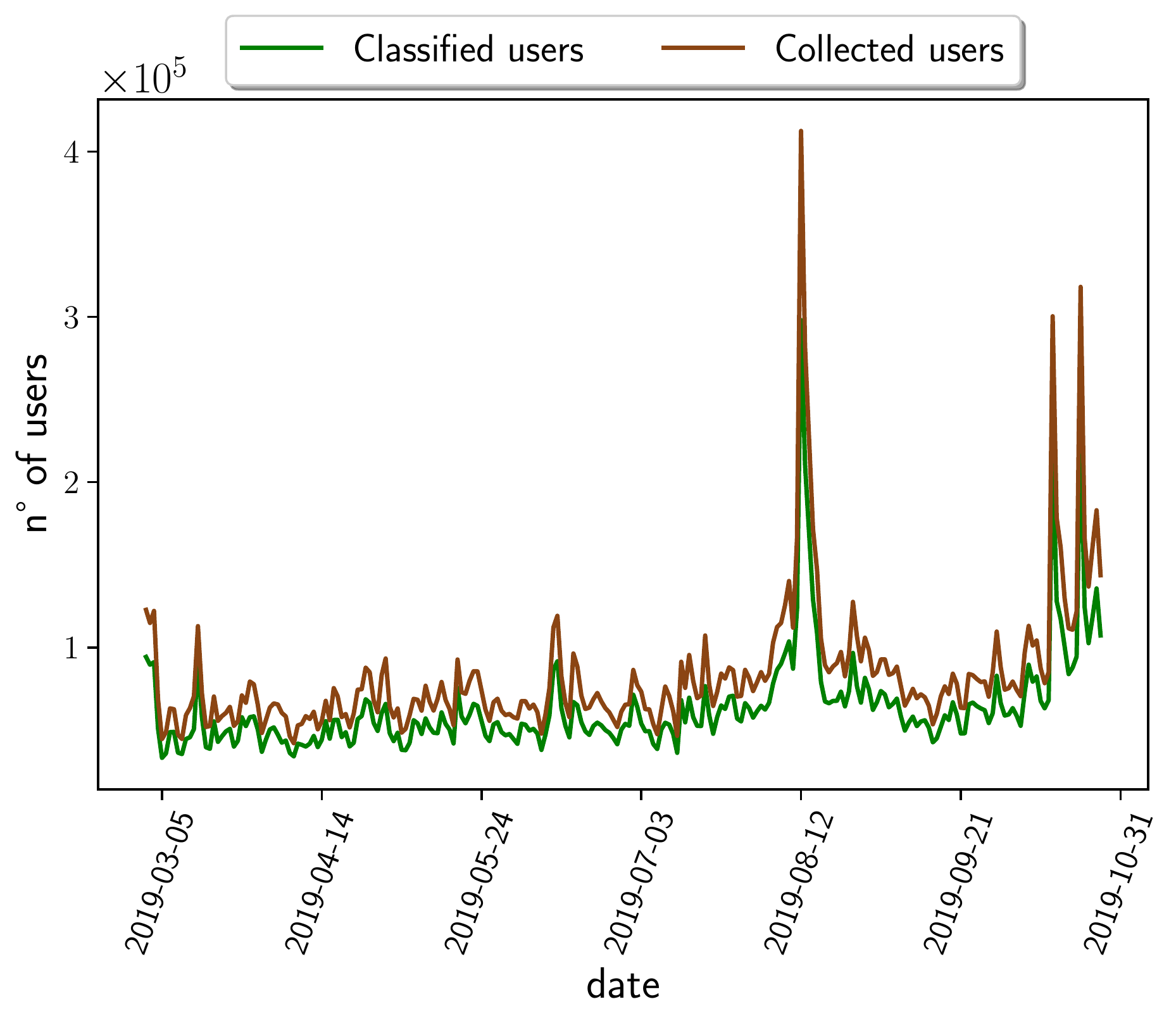}} \hfil
  \subfloat[]{\includegraphics[width=.5\textwidth]{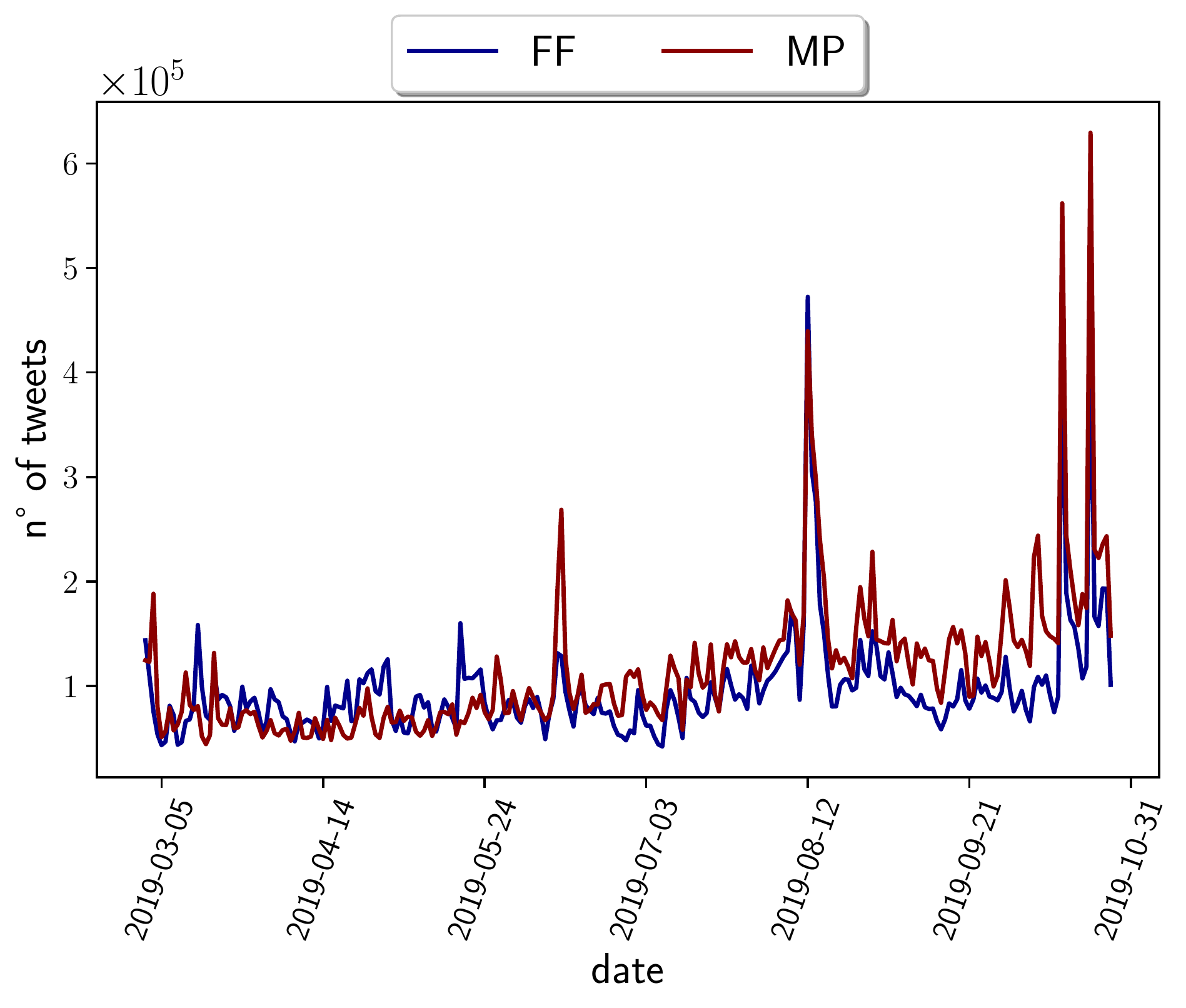}} \hfil
  \subfloat[]{\includegraphics[width=.5\textwidth]{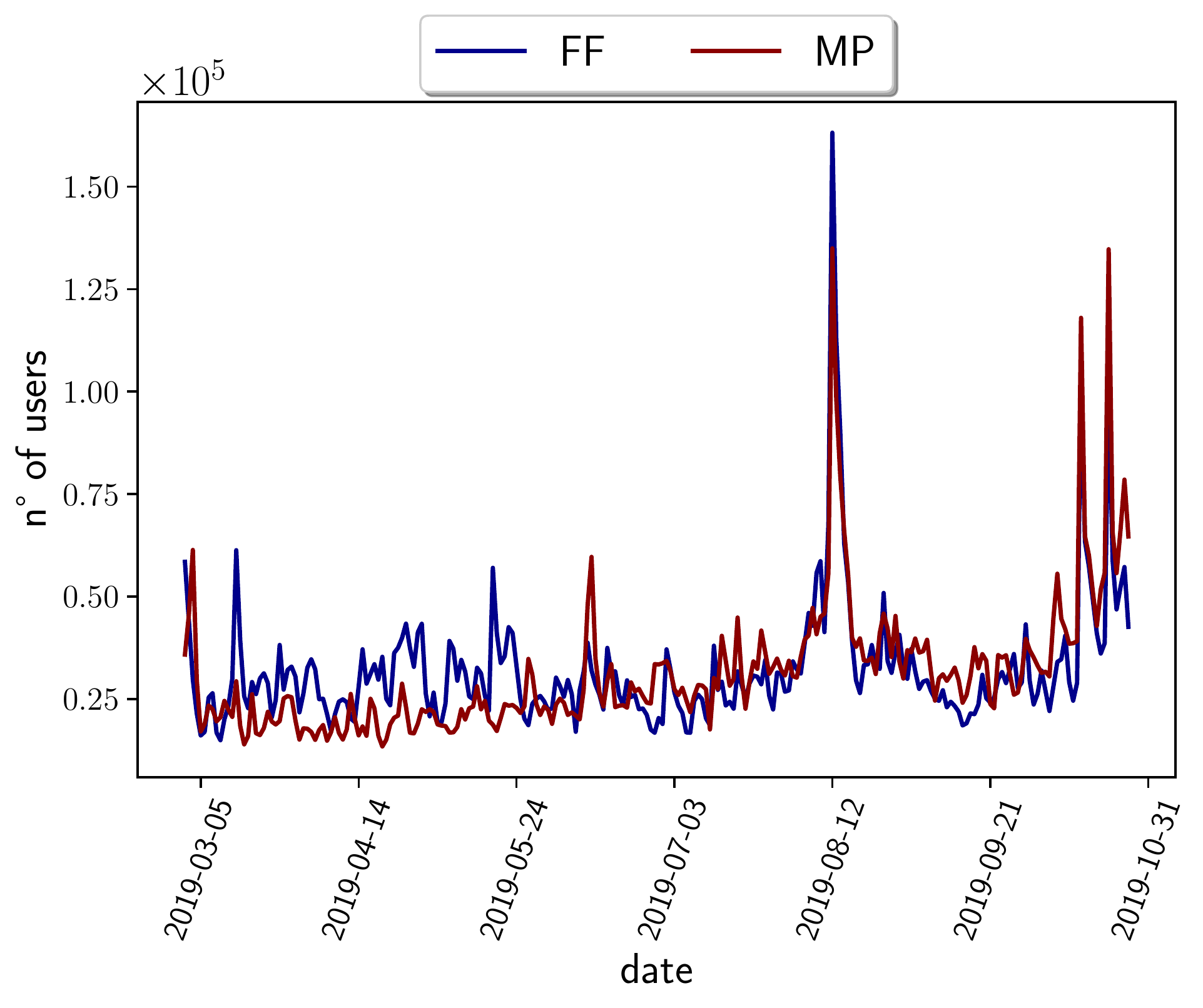}} \hfil
  \caption {(a) Daily volume of collected (brown line) and classified (green line) tweets. (b) Daily volume of collected (brown line) and classified (green line) users.
  (c) Daily tweets supporting the FF/MP formula. (d) Daily users supporting the FF/MP formula.}
 \label{fig:volume}
\end{figure}

\textbf{Data collection.} By means of the Twitter public APIs, we
collected tweets from March 1, 2019 until October 27, 2019, filtered
according to the following queries (corresponding to the candidates'
name and handlers of the 2019 Argentina primary election): {\it
  Alberto AND Fernandez}, {\it alferdez}, {\it CFK}, {\it
  CFKArgentina}, {\it Kirchner}, {\it mauriciomacri}, {\it Macri},
{\it Pichetto}, {\it MiguelPichetto}, {\it Lavagna}. Only tweets in
Spanish were selected. Figure  \ref{fig:volume}a shows the daily
volume of tweets collected (brown line) while Fig. \ref{fig:volume}b
shows the daily number of users (green line). In blue we report the
daily number of tweets/users which are classified, i.e. they posted at
least one classified tweet. Users are classified with machine learning
as supporters of Macri (Fig. \ref{fig:volume}d, red line) if the
majority of their daily tweets are classified in favor of Macri
(Fig. \ref{fig:volume}c, red line) or as supporters of Fern\'andez in
the other way around (blue line in Fig. \ref{fig:volume}d and
c). Hereafter we use FF to indicates the Fern\'andez-Fern\'andez
formula and with MP we refers to the Macri-Pichetto formula (the
outgoing president/vice-president candidate).

The activity of tweets/users shows a peak on August 11, 2020, i.e. the
day of the primary election. In the period from March to October, we
collected a daily average of 282,811 tweets posted by a daily average
of 84,062 unique users. We daily classified 75\% of these tweets and
$\sim$ 76\% of the users ( see Table \ref{tab:users_stat} and
Table \ref{tab:tweets_stat} in the \SI ). In total, by the end of
October we collected around 110 million tweets broadcasted by 6.3
million users. This large amount of tweets collected has no precedent
and is relevant in the light of considering that Argentina is one of
the most tweeting per capita countries in the world.

\textbf{User and text processing.} Below we explain the tasks that
need to be applied to the raw data before any analysis is performed.

\begin{figure}
  \centering
  \subfloat[]{\includegraphics[width=.5\textwidth]{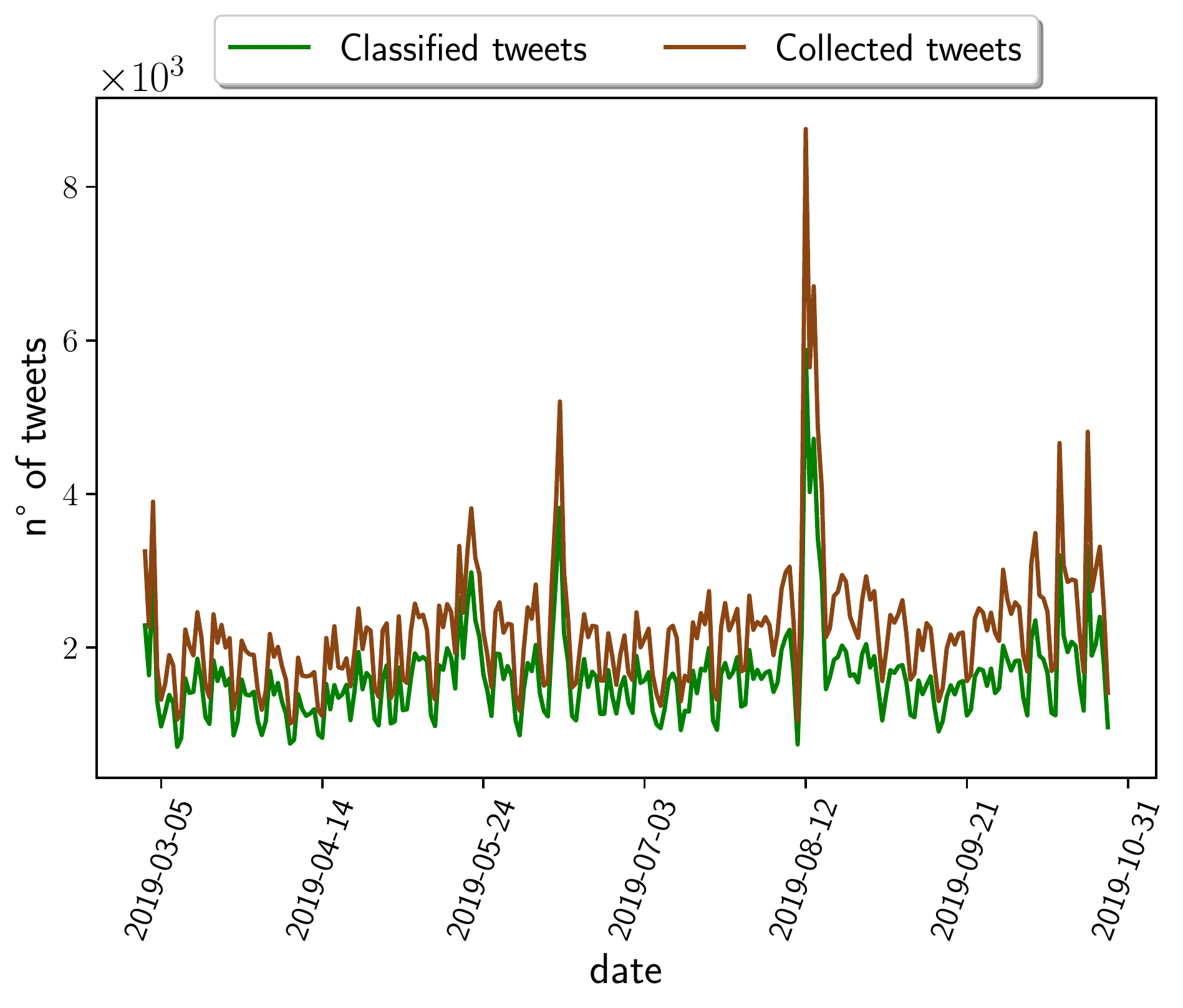}} \hfil
  \subfloat[]{\includegraphics[width=.5\textwidth]{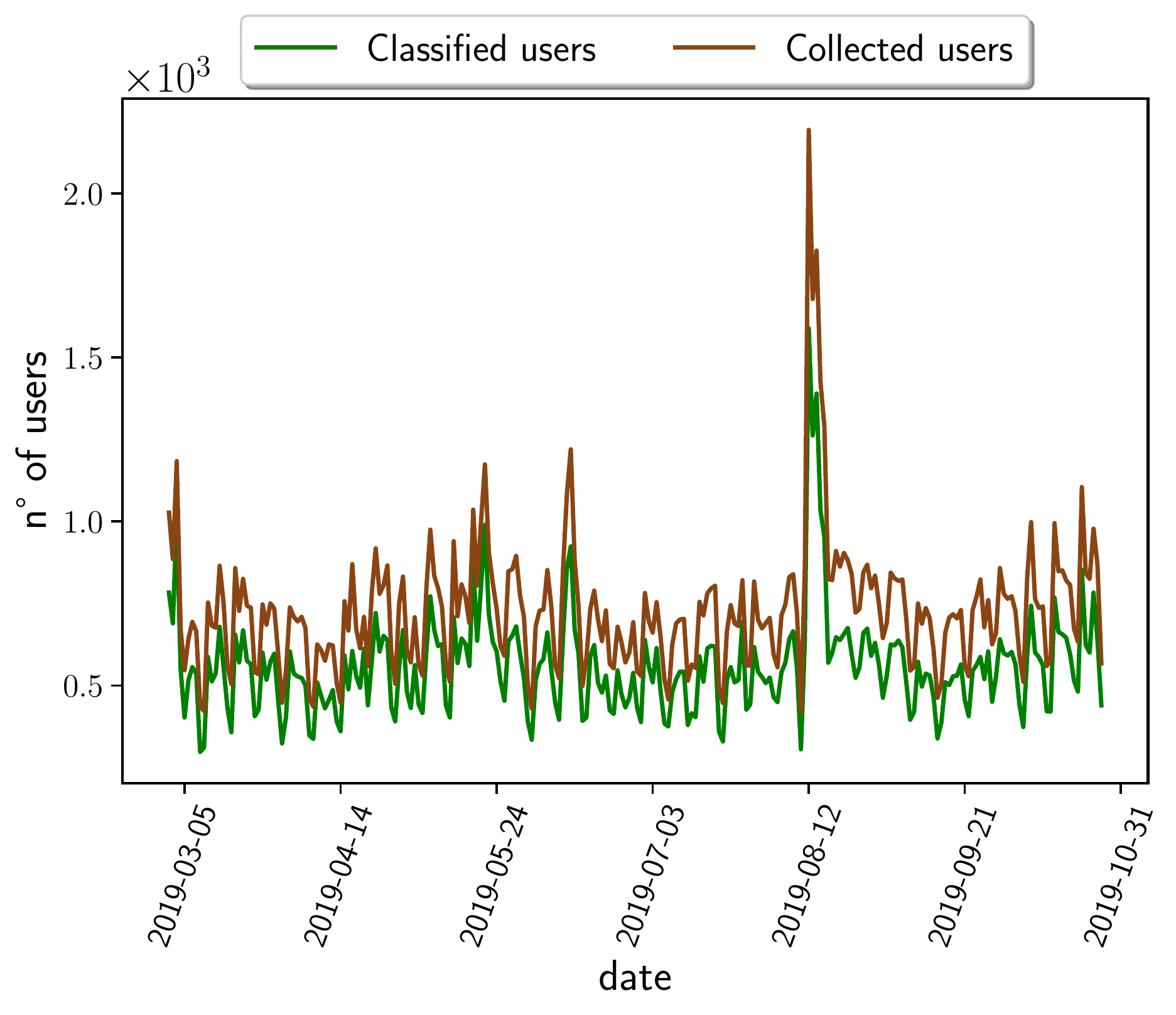}} \hfil
  \subfloat[]{\includegraphics[width=.5\textwidth]{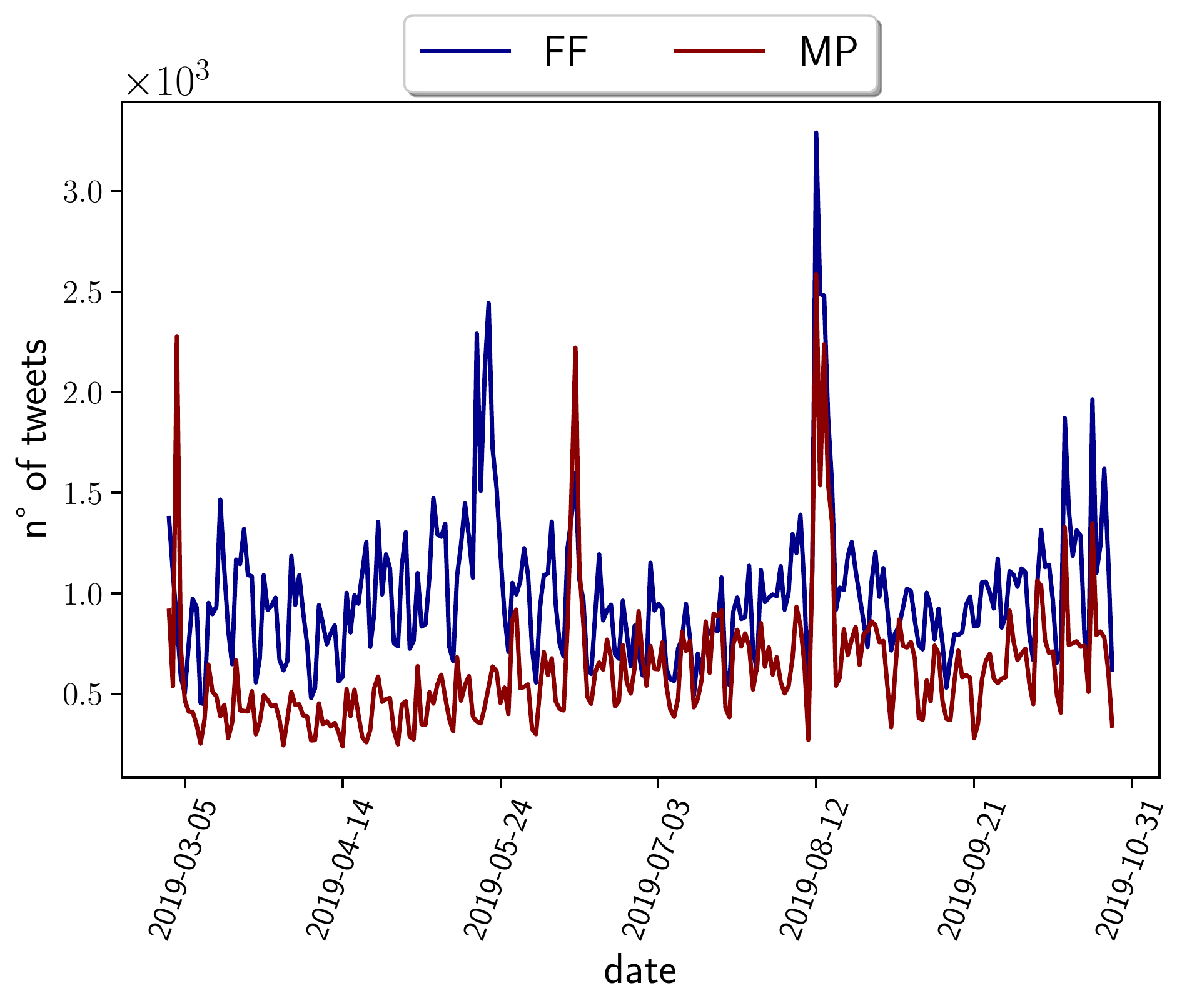}} \hfil
  \subfloat[]{\includegraphics[width=.5\textwidth]{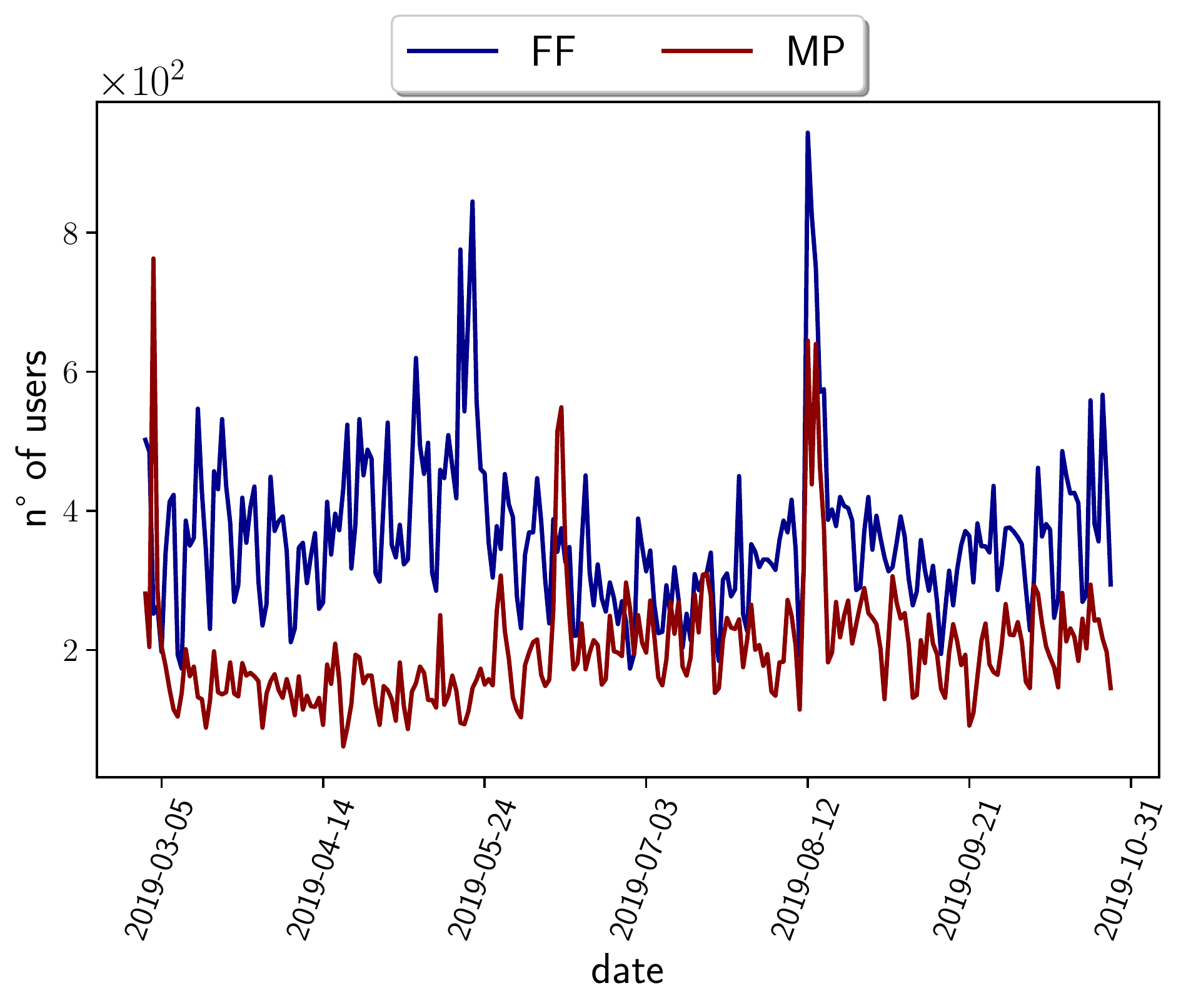}} \hfil
  \caption {Bots analysis: (a) Daily volume of collected (brown line) and classified (green line) tweets. (b) Daily volume of collected (brown line) and classified (green line) users.
  (c) Daily tweets supporting the FF/MP formula. (d) Daily users supporting the FF/MP formula.}
  \label{fig:bots}
\end{figure}

\textit{Bots detection.} The identification of software that
automatically injects information in the Twitter' system, is of
fundamental importance to discern between ``fake'' and ``genuine''
users \cite{KBS02}, the latter representing the real voters. According to
\cite{bovet} a good strategy is to extract the name of the Twitter
client used to post each tweet from their source field and kept only
tweets originating from an official Twitter client. 
Figure \ref{fig:bots}a and b show the daily number of tweets posted by bots
and the daily volume of bots, respectively. Figure \ref{fig:bots}c
and d show the daily volume of classified tweets/bots. The daily
average of bots between March and October is 732 with an overall daily
activity (in average) of 2,243 tweets. The daily classified tweets are
1,617 while the daily classified bots are 560 bots. As for ``genuine''
users, a bot is classified if it share at least 1 classified tweet. In
the entire dataset we found around 20,000 bots which posted 538,350
tweets. Let us notice that even though we classified the bots, they
are not used for the final prediction since they do not corresponds to
real voters.

\textit{Text standardization.} Stop words removal and word
tokenization are of common practice in Data mining and Natural
language processing (NLP) techniques \cite{NLP1,NLP2}. For example, we
keep the URLs as tokens since they usually point to resources
determining the opinion of the tweet, through replacing all URLs by
the token ``URL''.

\textbf{Tweets classification.} To build the training set we analyze
the hashtags in Twitter. Users continuously labels their tweet with
hashtags, which are acronyms able to directly transmit the user
feeling/opinion toward a topic. We hand labeled the top hashtags used
in the dataset (see Table~\ref{hashtag_table_March} in the \SI). They
are classified either as pro M(acri), F(ern\'andez) or T(hird party)
candidate, depending on who they support (with Third party we refer to
the supporters of Lavagna, Espert and other secondary candidates).

\begin{figure}[!ht]
   \includegraphics[width=.9\textwidth]{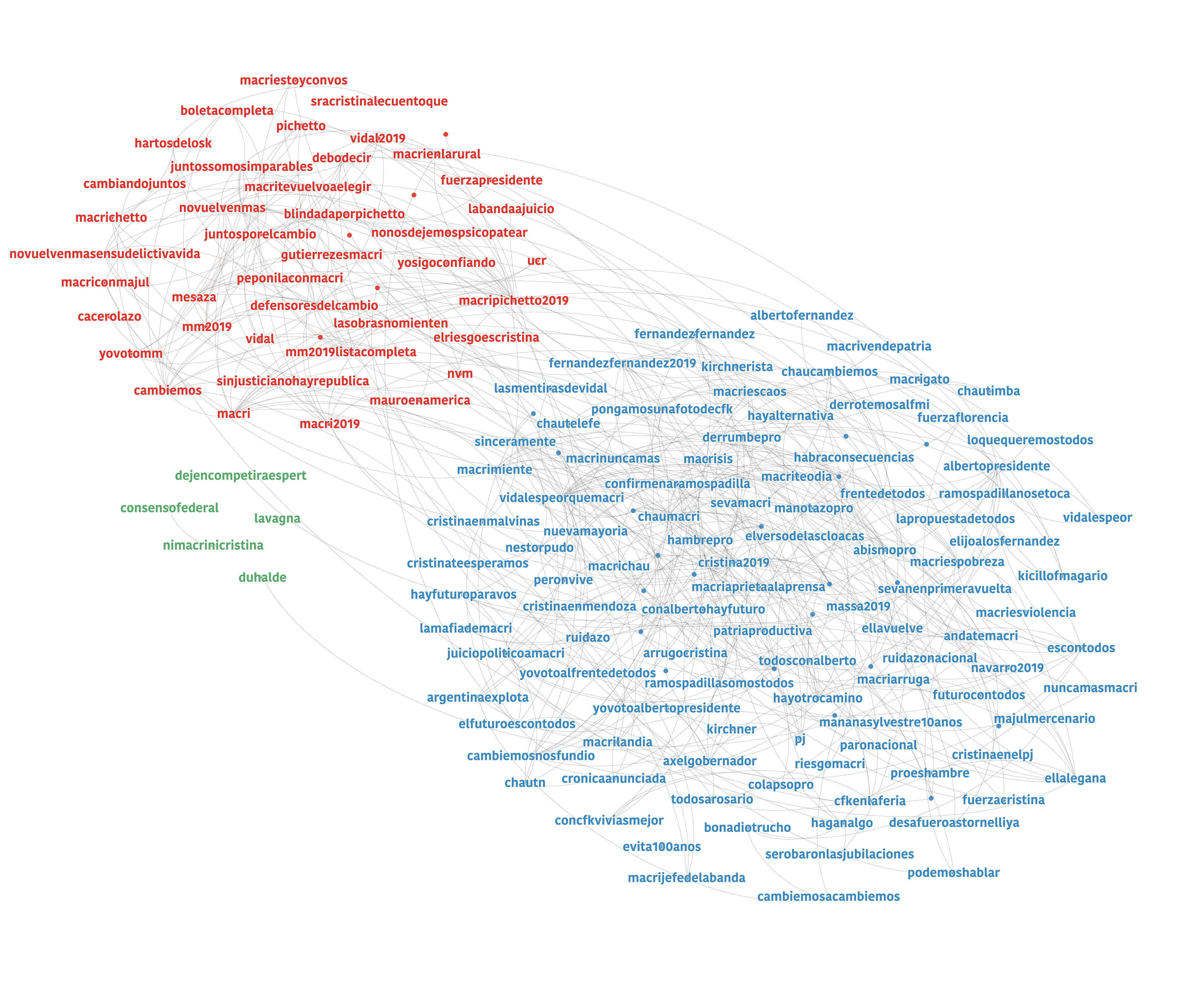}
   \caption {Hashtag co-occurrence network from March to August
     2019. In blue the hashtags in favor of Alberto Fern\'andez and
     Cristina Fern\'andez de Kirchner, in red the hashtags in favor of
     Macri and in green those in favor of the Third party.}
  \label{hashtags}
\end{figure}

\begin{figure}[!ht]
  \centerline{
  (a) \includegraphics[width=.5\textwidth]{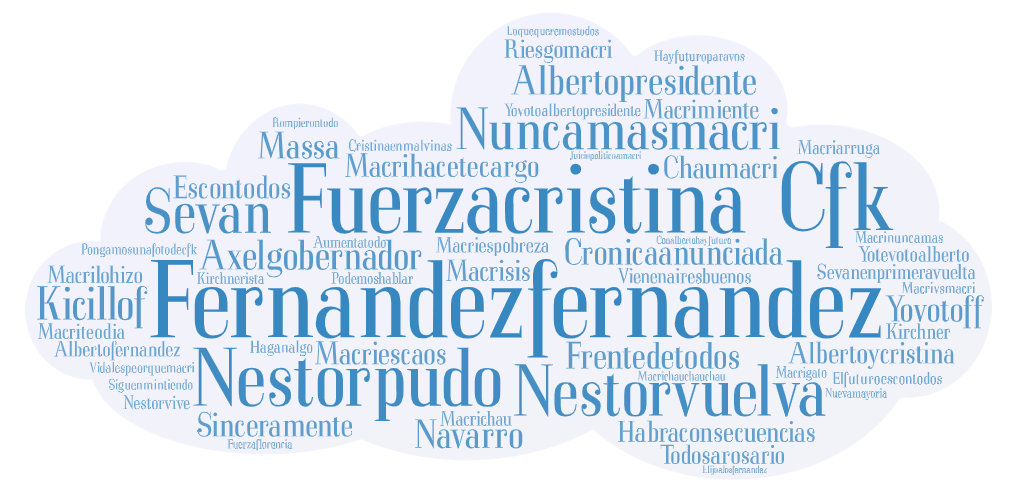}
  (b) \includegraphics[width=.5\textwidth]{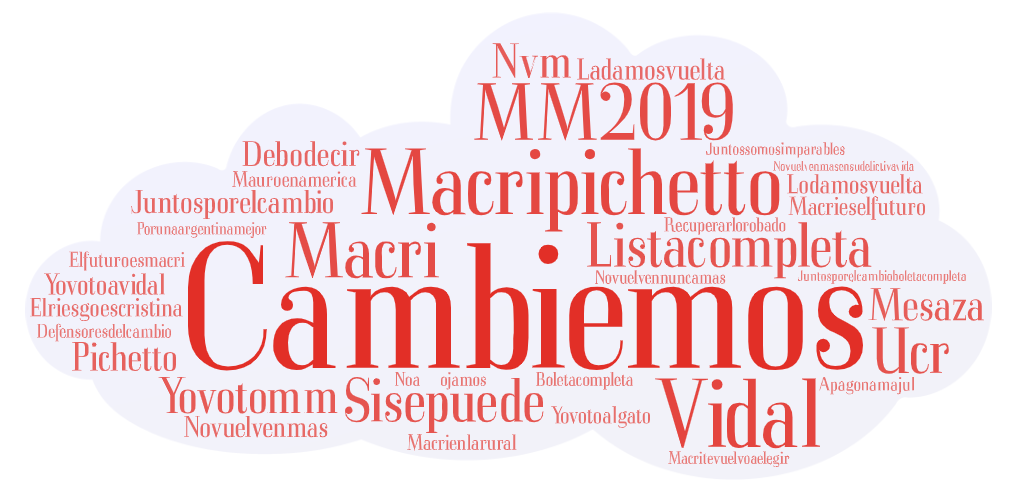}}
  \caption{Hashtag clouds. The dimension of the words is proportional to  
  their frequency in the dataset. (a) The blue hashtags are the most frequently used in the tweets in
  favor of Fern\'andez. (b) The red hashtags are those in favor of Macri.}
  \label{main_hashtags}
\end{figure}

\textit{Hashtag co-occurrence network.} In order to check the quality
of the classification of the classified hashtags we build the hashtag
co-occurrence network $H(V, E)$ and statistically validate its edges
\cite{martinez2011disentangling,bovet}. In the co-occurrence network
the set of vertices $v \in V$ represents hashtags, and an edge
$e_{ij}$ is drawn between $v_i$ and $v_j$ if they appear together in a
tweet. We test the statistical significance of each edge $e_{ij}$ by
computing the probability $p_{ij}$ (p-value of the null hypothesis) to
observe the corresponding number of co-occurrences by chance only
knowing the number of occurrences $c_i$ and $c_j$ of the vertices
$v_{i}$ and $v_{j}$, and the total number of tweets
$N$. Fig.~\ref{hashtags} shows the validated network. We only keep
those edges with a p-value $p<10^{-7}$. The blue community contains
the hashtags in favor of Fern\'andez, the red community those in favor
of Macri and the green one (a very small group) are those in favor of
the Third candidate. A look at the typologies of hashtags reveals the
first differences in the supporters. Those in favor of Cristina
Kirchner are much more passionate than the follower of Macri. For
example, Kirchner's type of hashtags are \#FuerzaCristina,
\#Nestorvuelva, \#Nestorpudo or they are very negative to Macri as
\#NuncamasMacri. On the other hand, Macri's group is smaller and less
passionate with hashtags like \#Cambiemos or \#MM2019 (see
Fig. \ref{main_hashtags}), while support for the third candidate has
not taken traction and its electoral base on Twitter is very small.

In principle, counting the users and tweets according to the hashtags
they use would predict the victory of Fern\'andez over Macri. However
this conclusion would be based only on $\sim$ 10,000 users (those
expressing their opinion through hashtags). In order to get the
opinion of all the users we train a machine learning model that
classifies each tweet as AF, MM or Third party. (In what follows we
also refer to the formulas FF for Fern\'andez-Fern\'andez and MP for
Macri-Pichetto, the final formulas in the presidential contest).  We
use the previous set of hashtags expressing opinion to build a set of
labeled tweets, which are used in turn to train a machine learning
classifier. We use all the tweets (before August) which contain at
least one of the classified hashtags to train the model. In the case
of more than one hashtag for a tweet, we consider it only if all the
hashtags are in favor of the same candidate. The use of hashtags that
explicitly express an opinion in a tweet represents a ``cost'' in
terms of self-exposition by Twitter users \cite{Ceron} and therefore
allows one to select tweets that clearly state support or opposition
to the candidates. The training set consists of 228,133 tweets,
i.e. the 0.33\% of the total amount of collected tweets and the
$\sim$90\% of the hand-classified tweets (253,482 tweets). In order to
find the best classifier we used five different classification models,
the logistic regression (LR) with $L_2$ regularization, the support
vector machine model (SVM), the Naive Bayes method (NB), the Random
Forest (RF) and the Decision Tree (DT). All these models are validated
on the remaining 10\% of the classified tweets (25,349). Table
\ref{tab:accuracy} shows the results for the models. The logistic
regression performs better than the other models with an average group
accuracy equals to 83\%. Also recall and F1-score are equal to
83\%. Support Vector Machine is the second classified, with an average
accuracy of 81\%. It follows the Naive Bayes with and average accuracy
of 79.5\%, the Random Forest and the Decision Tree.

\begin{figure}[!ht]
  \includegraphics[width=.6\textwidth]{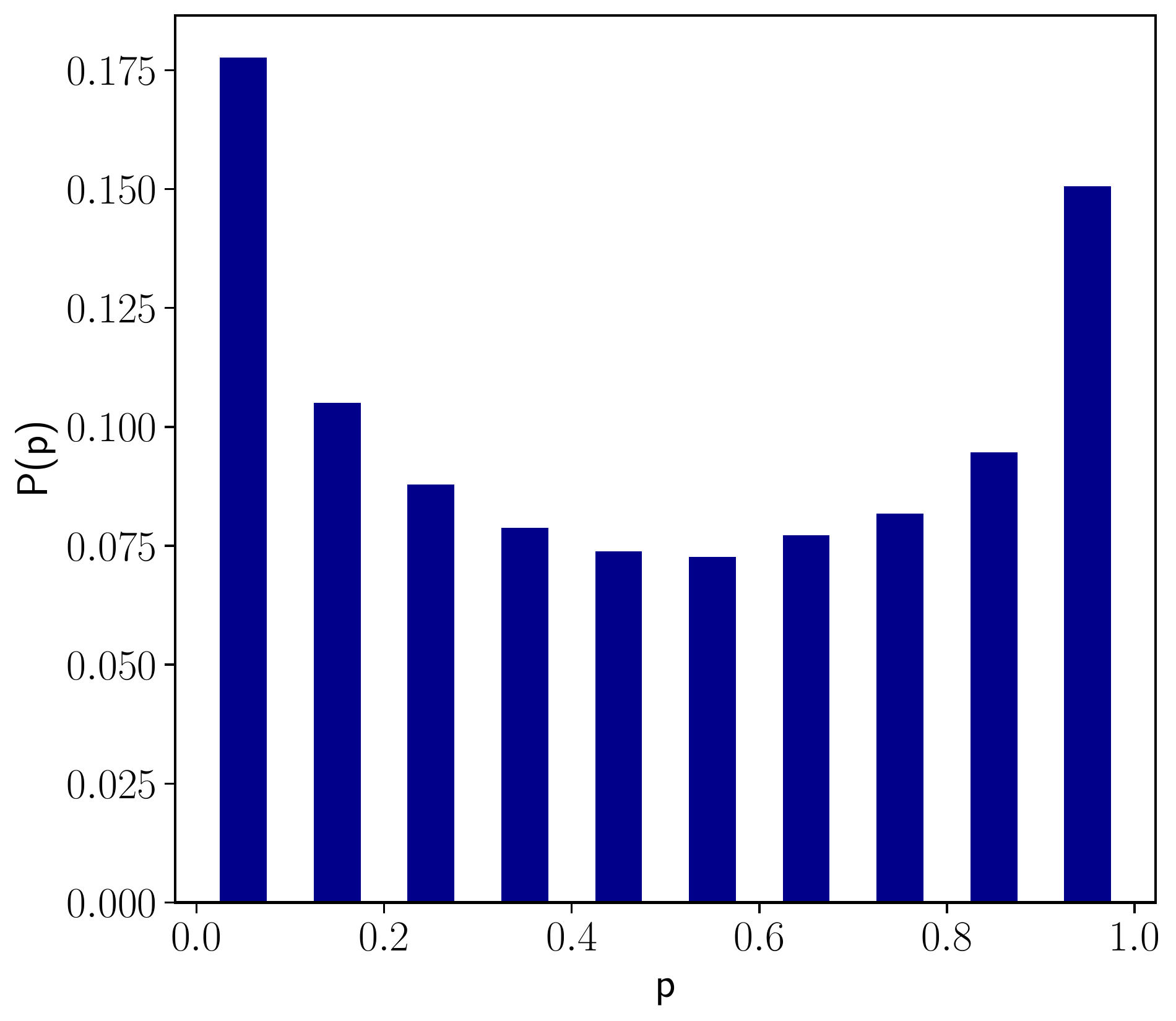}
  \caption {Probability distribution of $p$ = probability of voting
    for Macri ($p=0$ corresponds to Fern\'andez) obtained by the
    Logistic Regression model.}
  \label{fig:probablity}
\end{figure}

We recall that the logistic regression assigns to each tweet a
probability $p$ of belonging to a class. In our case such probability
goes to one if the tweet supports Macri while it goes zero if it
supports Fern\'andez. As it is shown in Fig. \ref{fig:probablity} the
distribution of $p$ contains two peaks, one on the left and one on the
right, divided by a plateau. This is an encouraging result, since it
proofs the efficacy of the model to discern between the two
classes. We classify a tweet in favor of Macri if $p \geq 0.66$, in
favor of Fern\'andez if $p \leq 0.33$. Tweets with a value of $p$ in
the plateau are instead unclassified, meaning that the tweet does not
contain sufficient information to be classified in either
camp. According to this rule, in average we classify 211,229 genuine
tweets and 1,617 ``fake'' tweets per day (see
Table~\ref{tab:users_stat} and Table~\ref{tab:tweets_stat} in the
\SI).

\textbf{Opinion modeling.} We can infer users' opinion from the
majority of the tweets they post. Let $n_{t,F}$ be the number of
tweets posted by a given user at time $t$ in favor of Fern\'andez and
let $n_{t,M}$ be those supporting Macri. We define an instantaneous
opinion over a window of length $w$ and a cumulative average opinion
as follow. In the first case, a user is classified as a supporter of
Fern\'andez (at a given day $t=d$) if $\sum_{t=d-w+1}^{d}n_{t,F} >
\sum_{t=d-w+1}^{d}n_{t,M}$, i.e if the majority of the tweets posted
in the last $w$ days were in favor of Fern\'andez. The user is
classified as a supporter of Macri if $\sum_{t=d-w+1}^{d}n_{t,F} <
\sum_{t=d-w+1}^{d}n_{t,M}$. If none of the previous conditions is met,
i.e. if $\sum_{t=d-w+1}^{d}n_{t,F} = \sum_{t=d-w+1}^{d}n_{t,M}$ then
the user is classified as undecided. Let us notice that when $w$ goes
to one we have the `most' instantaneous prediction, that is the
prediction based on what people think in the last day. This
instantaneous prediction model was used in Ref. \cite{bovet} to match
the results of the AI model to the aggregate of polls from the New
York Times in the 2016 US election with excellent results. However,
this predictor did not match the results of the electoral college,
which required stratification by states. Thus, we further develop the
AI model of \cite{bovet} to add other predictors beyond the
instantaneous measures.

Traditional polls' data collection is an instantaneous prediction with
a value of $w$ that can go from few days up to few weeks, which is the
time of collection of the poll data and this corresponds roughly to
our instantaneous measurement above.  However, the fact that we are
able to track the same user over long period of time in Twitter allows
us to extend the window of observation as far as we want to then
define a new measure that we call the cumulative opinion.  The
cumulative opinion in our model is defined by extending $w$ to the
initial date of collection for every time $d$ of observation, i.e.,
$w=d$, thus considering the opinion of a user based on all the tweets
he/she posted from time $t=0$ upto the observation time $d$. That is,
our prediction is longitudinal as we are able to follow the opinion of
the same user over the entire period of observation of several months.
In terms of traditional poll methods, a cumulative opinion would be
obtained in a panel collecting for each respondent in the sample and
for each day starting from $t=0$ her/his preference toward a
candidate.  This possibility, which would require an unimaginable
amount of effort and time for traditional poll methods, it is quite
straightforward when it come to social media and big-data analyses.

\begin{figure}
  \centering
  \subfloat[]{\includegraphics[width=.60\textwidth]{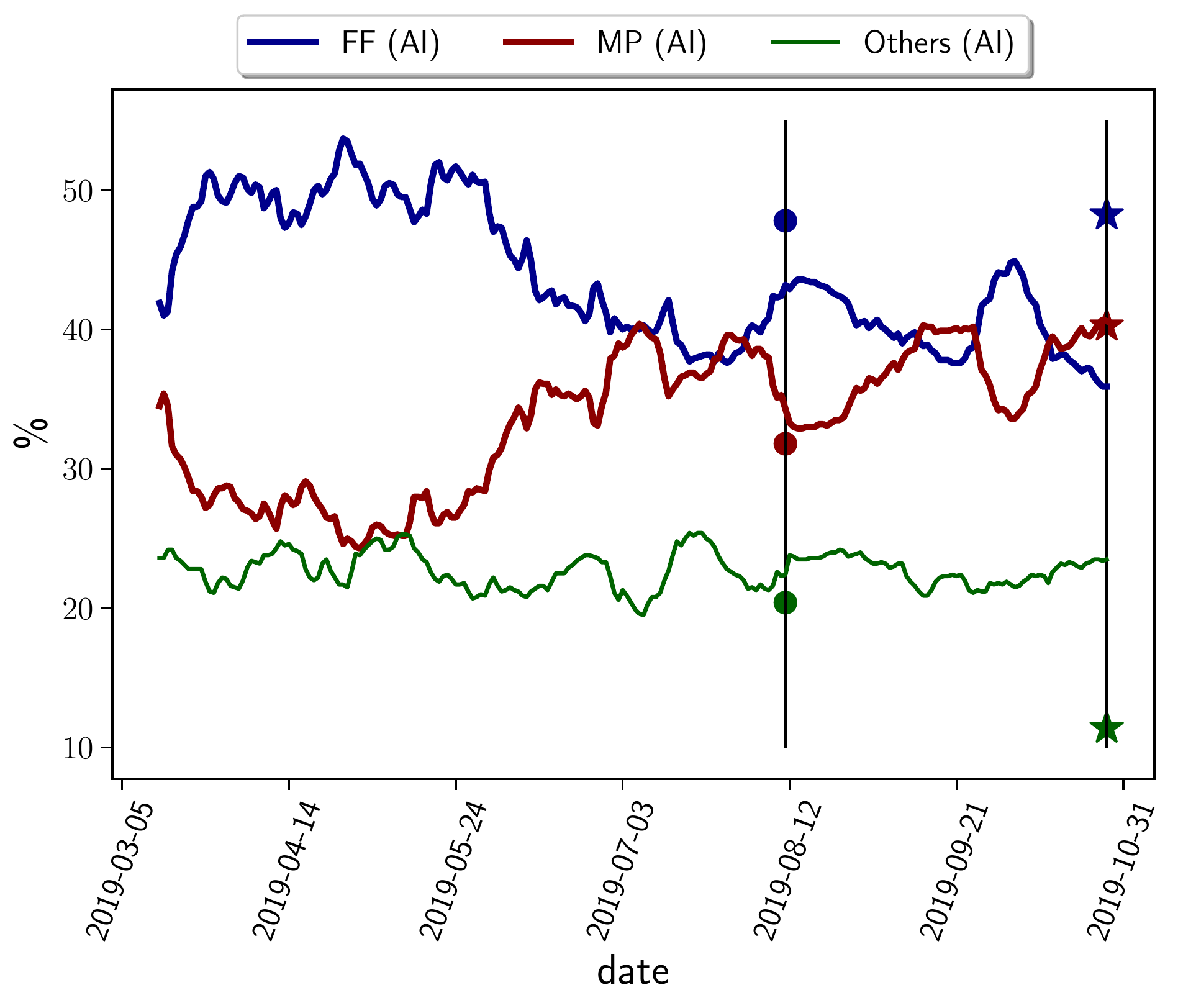}} \hfil
  \subfloat[]{\includegraphics[width=.60\textwidth]{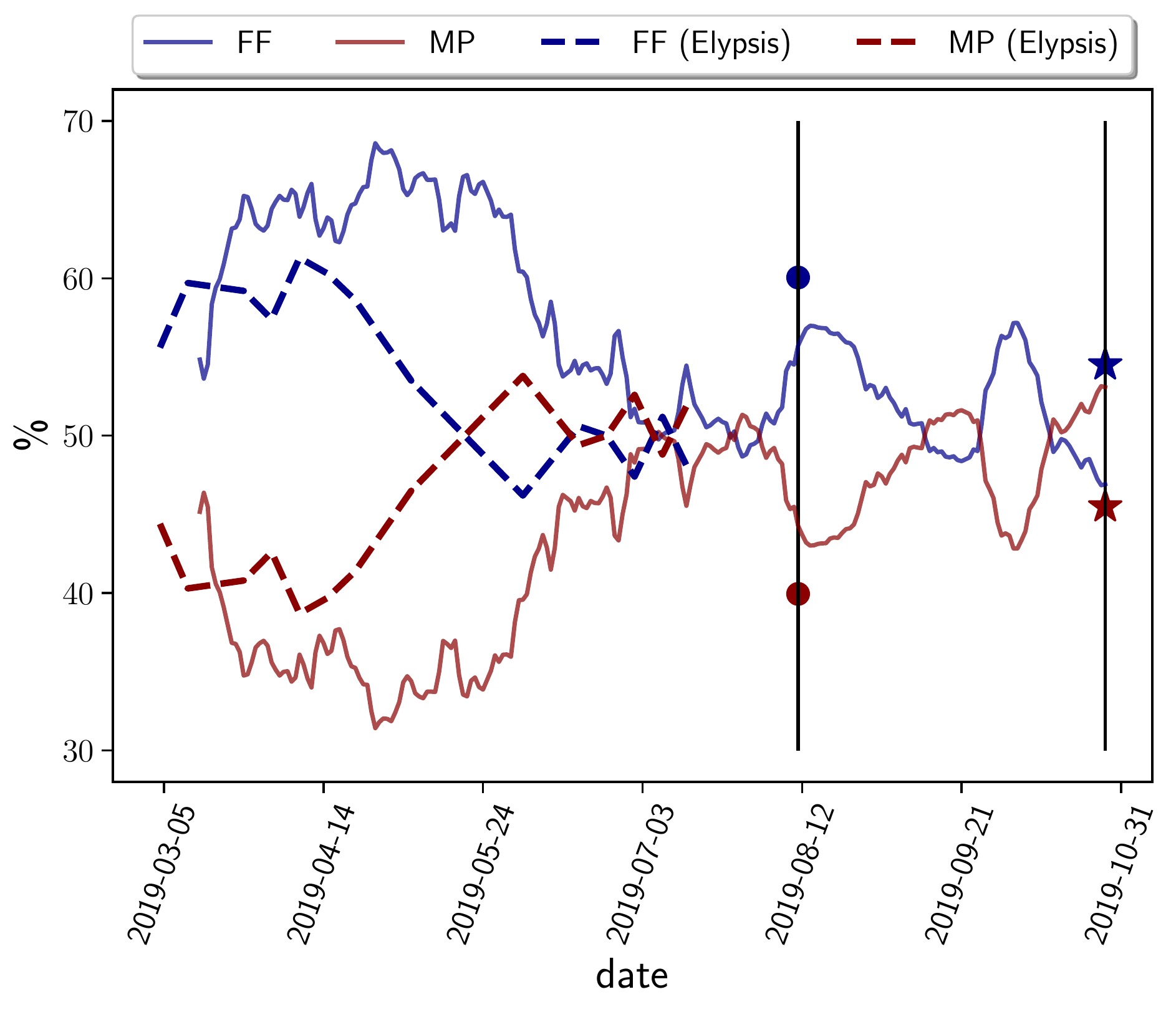}}
  \caption{(a) Instantaneous prediction of the AI model obtained in a moving window of $w$=14 days. Vertical lines 
  are ( from the left to the right ) the day of the primaries and general elections respectively. The circles represents the official results for the primaries while the stars those for the general elections. (b) Previous results compared with
    polls from Elypsis. Thick lines represent AI prediction while
    dashed line represent the Elypsis predictions.}
  \label{fig:14days}
\end{figure}
\begin{figure}
  \centering
  \includegraphics[width=.9\textwidth]{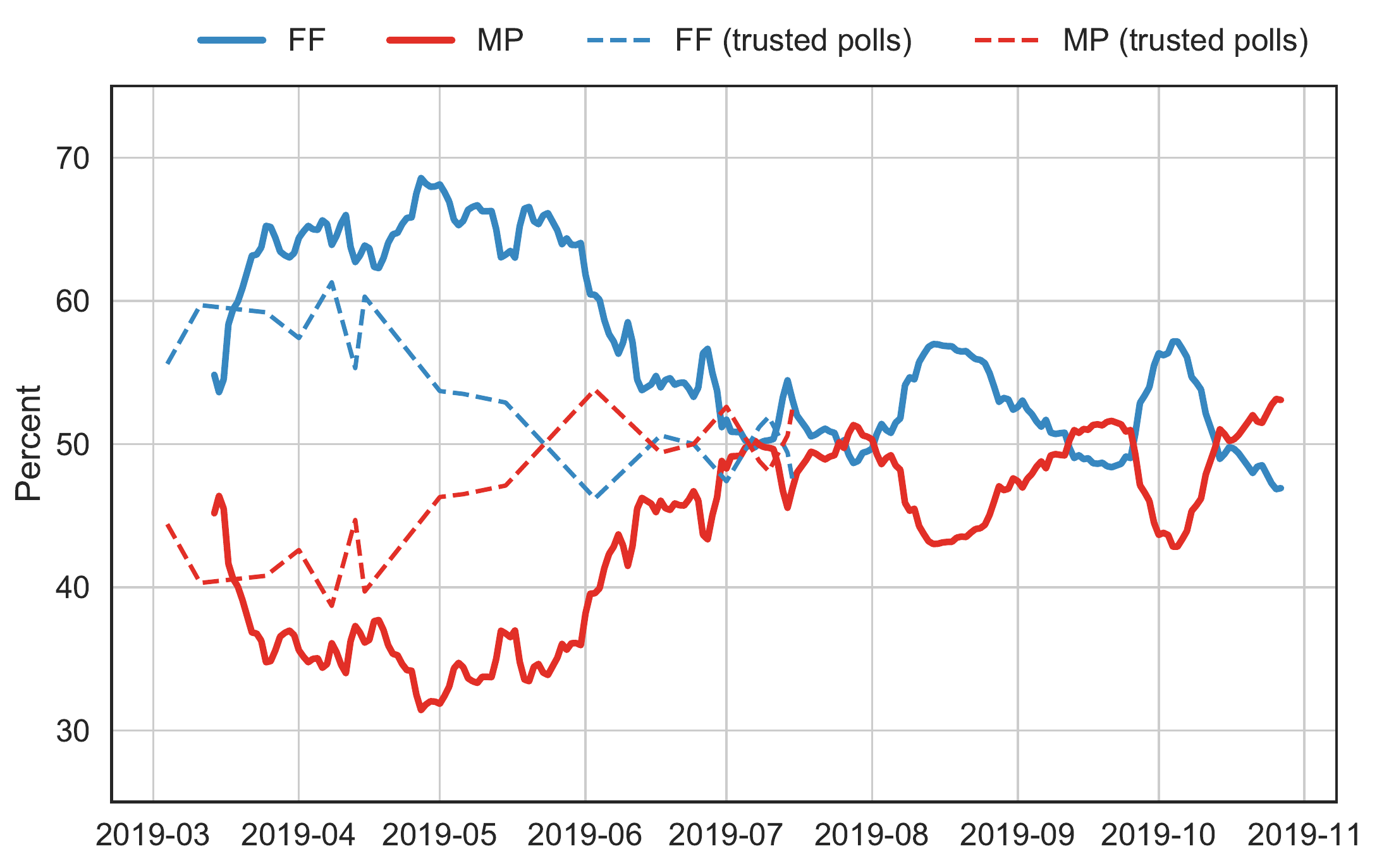}
  \caption{Instantaneous prediction compared with trusted polls. Thick lines represent AI prediction and dashed line represent trusted polls.}
  \label{fig:14days_trusted_polls}
\end{figure}

We start by investigating the instantaneous response of the users in a
fixed window of time. Figure \ref{fig:14days}a shows the Twitter
supporters dynamics over time obtained with a window average, $w=14$
days. Users are classified as MP (in red), FF (in blue) or Others (in
green). Figure \ref{fig:14days}b shows the supporters dynamic (thick
lines) compared with Elypsis prediction (thin dashed lines) without
considering the undecided users in the normalization. In the same plots we also report the official results for both primaries and general elections. The comparison
between the two pictures stands out as a approximate correlation
between the Elypsis and the AI results for each candidate. However, in
the comparison among candidates predictions may sometimes differs, as
for example, right before the beginning of August, Elypsis gave as
favorite MP while the AI instantaneous prediction was in favor of
FF. Overall, as for the pollsters results, window average analyses are
representative of the instantaneous sentiment of the people. As we see
from the figures, instantaneous opinions are affected by considerable
fluctuations \cite{bovet} which make the prediction not reliable. In
Fig. \ref{fig:14days_trusted_polls} in the \SI we compare the average
window opinion with other pollsters (Real Time Data, Management \&
Fit, Opinaia, Giacobbe and Elypsis). An interpolation (thin lines)
shows similar trends as the AI-model window average, stressing that
the conclusions made so far are more general then the simple
comparison with Elypsis. In fact in \cite{bovet} we have shown that
the instantaneous predictions of the AI model follows quite closely
the aggregation of polls obtained from the New York Times, `The
Upshot', yet, it does not reproduce the results of the electoral
college which requires a segmentation by states where proper
prediction of rural and non-rural areas becomes the key and
considering the cumulative opinion, not the instantaneous one, opinion
of each users is crucial to correctly predict the elections.


Thus, we next study the opinion of each user by considering the
cumulative number of tweets over the entire period of observation to
define classify the voter's intention (\textsc{Model 0}). This cumulative approach takes
into consideration the all the tweets together for each user since the
first time they enter in the dataset and based the voter intention on
all of them.  This cumulative approach can only be done with Twitter
and not with traditional polls, except for short times and particular
cases as done by Elypsis before and after PASO.

\begin{figure}[!ht]
  \includegraphics[width=.65\textwidth]{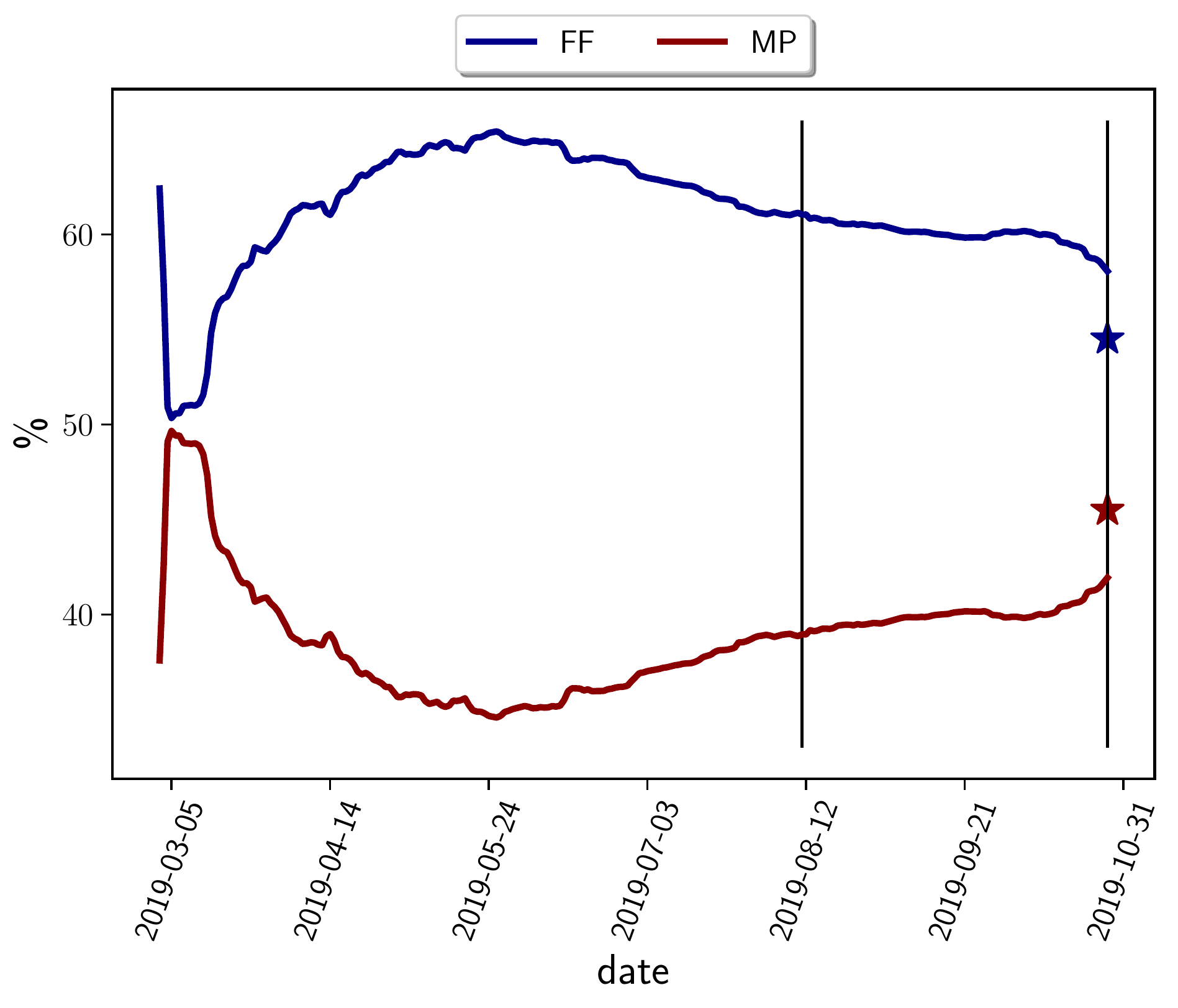}
  \caption{Simple cumulative predictions obtained without defining the loyalty classes (\textsc{Model 0}).}
\label{fig:M1}
\end{figure}

Figure \ref{fig:M1} shows the cumulative opinion from March 1 until a
few days before the general elections. We can see that this approach captures the
huge gap between the candidates, both for the primary election and the
general election ( vertical lines from the left to the right). While a low precision is of secondary importance
when the difference between the opponents is high, it plays a central
role when they have a close share of supporters. As an extreme
example, in an almost perfect balanced situation the change of mind of
just few people may flip the final outcome. If on the one hand a
cumulative approach do reduce the fluctuations in the signal, it is
also less sensitive to sudden change of opinion. A person can support
a candidate until few days before the elections, for then change
her/his mind because of some particular facts. This and other
possibilities can be taken into account only by a model based on
cumulative analyses, but able to capture the degree of loyalty of
people towards the candidates over time. Differently from the
traditional surveys, the real time data processing that underlies our
AI algorithm gives the possibility to take into consideration this
scenario.  To understand how different re-weighting scenarios affect
the results, below we introduce different loyalty classes of users
towards the candidates and then we define several models matching the
criteria previously discussed. These loyalty classes can be only
defined when we consider the cumulative opinion in a longitudinal
study and cannot be investigated by traditional polls.

\textit{Loyalty classes.} We define 5 classes of loyalty for
users. Here we consider the MP supporters, but the definitions below
similarly applied to the other candidates.

\begin{itemize}
\item Ultra Loyal (UL): users who always tweet only for the same candidate, namely $\sum_{t=T_0}^T (\frac{n_{M,t}}{n_{M,t}+n_{F,t}+n_{T,t}}) = 1$. where with $n_{x,t}$ we indicate the number of tweets that a given user post in favor of $x$, with $ x\in$ $\lbrace$ Macri, Fern\'andez, Third party$\rbrace$. 
\end{itemize}

Differently from the ultra loyal, which continuously post in favor of a candidate, the other classes take into consideration a possible change of opinion of a user. In order to detect sudden twist of opinions we focus on the classifications of the last $k$ tweets posted by the users. We define:

\begin{itemize}
\item Loyal MP $\to$ MP: a user which is MP since the majority of tweet are for MP, but she/he also supported MP in the last $k$ tweets. Mathematically speaking $\sum_{n=N-k}^N n_{M,n} > \sum_{n=N-k}^N n_{F,n}+n_{T,n}$. $N$ is the total number of tweets posted by the user.

\item Loyal MP $\to$ FF: users that are MP by the total cumulative count but they have tweeted for FF in the recent $k$ tweets. In formula: $\sum_{n=N-k}^N n_{F,n} > \sum_{n=N-k}^N n_{M,n}+n_{T,n}$

\item Loyal MP $\to$ TP: users supporting the third party in the last $k$ tweets, 
i.e. $\sum_{n=N-k}^N n_{T,n} > \sum_{n=N-k}^N n_{M,n} +n_{F,n}$.

\item Loyal MP $\to$ Undecided: all other individuals classified as MP but not included above.
\end{itemize} 

\begin{figure}[!ht]
  \includegraphics[width=.9\textwidth]{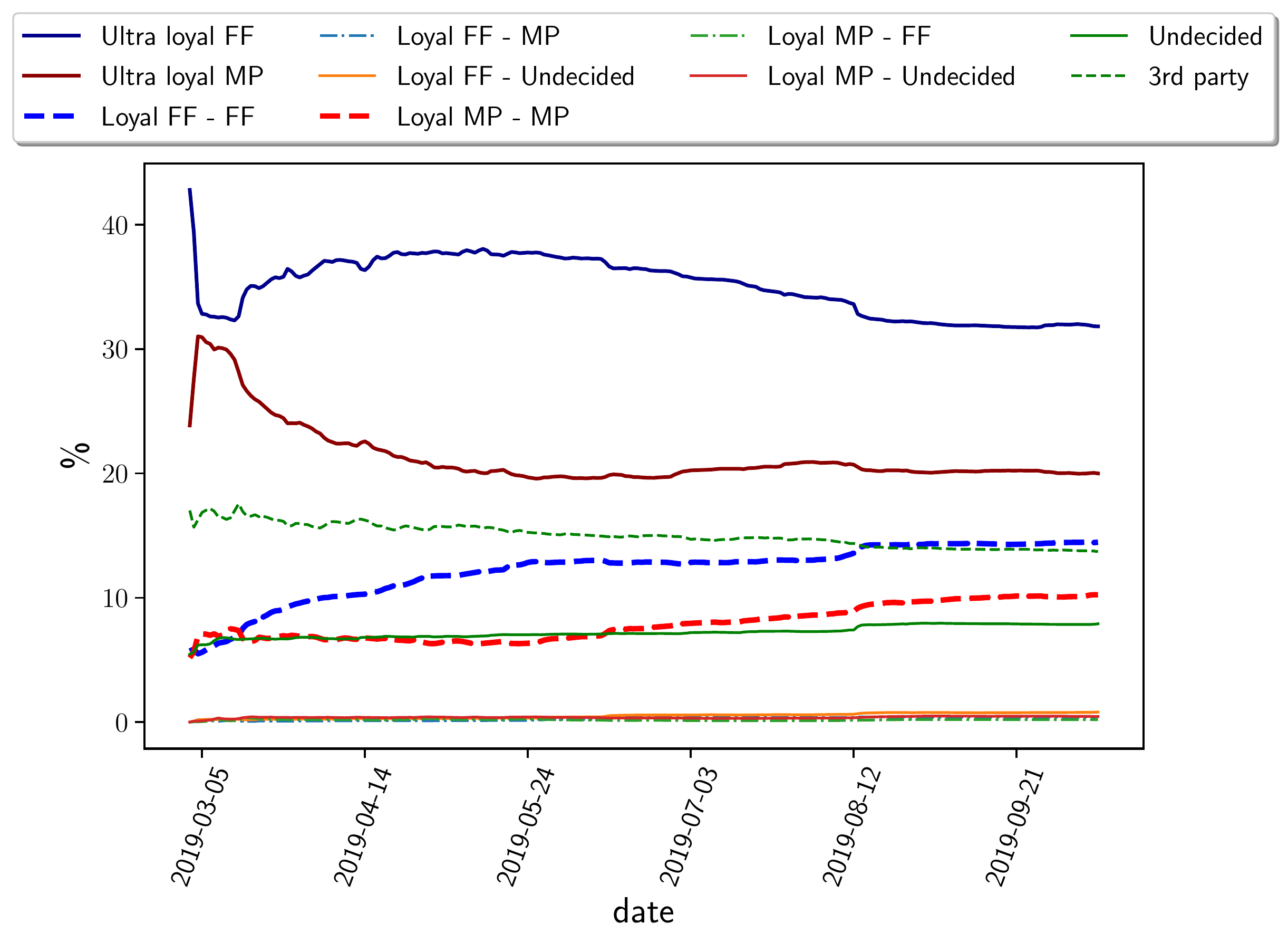}
  \caption{Cumulative users' opinion for each loyalty class over time.}
  \label{lclasses}
\end{figure}

Let us remind that unclassified refers to all those users who do not
have any classified tweet. Fig.~\ref{lclasses} shows the cumulative
prediction for each class, with $T_0$= March 1, 2019 and $k=10$. The
Ultra Loyal class for Fern\'andez (FF) represents $\sim$ 33\% of the
populations while only $\sim$ 20\% of the populations is Ultra Loyal
towards Macri (MP). Loyal MP$\to$MP and loyal FF$\to$FF represents
between the 8\% and the 13\% of the entire Twitter population. The
percentage of the undecided is around 8\% and the third party
percentage. The other classes are close to 1 or 2\%. In the next
section we use these classes in order to define a better predictor.

\textit{AI Models.} The loyalty classes introduced so far are one of
the main differences with the other Twitter based studies: we use the
machine learning classifier (logistic regression here) to define the
loyalty of a user and not to make predictions. We do that by grouping
supporters as follows:
\begin{itemize}
	\item Fern\'andez supporters: all those users which are 
	ultra loyal FF, loyal FF$\to$FF, loyal FF$\to$MP, loyal FF $\to$ Undecided.
	\item Macri supporters: all those users which are ultra loyal MP,
	 loyal MP$\to$MP, loyal MP$\to$FF, loyal MP$\to$Undecided.
\end{itemize}

In each group we put those users we are almost sure who they support
because of their activity over time. However, as we saw in the
previous section, undecided may play a central role in a scenario
where few percentage points can flip the final result. Furthermore,
understanding unclassified users (i.e. those users which do no not
have any classified tweet) will also improve the final statistic. In
order to take into account all the reasonable scenario we define three
different models (starting from the classification in Fern\'andez and
Macri of above) and validate them against the final results of the
election. Table \ref{tab:models} resumes the details of each model.

\textsc{Model 1:} All the users belonging to one of the following
classes are grouped in the third party: Undecided$\to$MP,
Undecided$\to$FF, Undecided$\to$Undecided and Unclassified.

\textsc{Model 2:} Instead of simply grouping the undecided in a third
party, we use network homophily to infer their political
orientation. A user is classified as MP(Undecided) if the majority of
her/his neighbors (in the indirected retweet network) support
Macri. The same definitions applied for the other cases. In this
model, FF(Undecided) are considered supporters of Fern\'andez and
MP(Undecided) supporters of Macri. Undecided(Undecided) and
Unclassified belong to the third party.

\begin{figure}[!ht]
  \includegraphics[width=.9\textwidth]{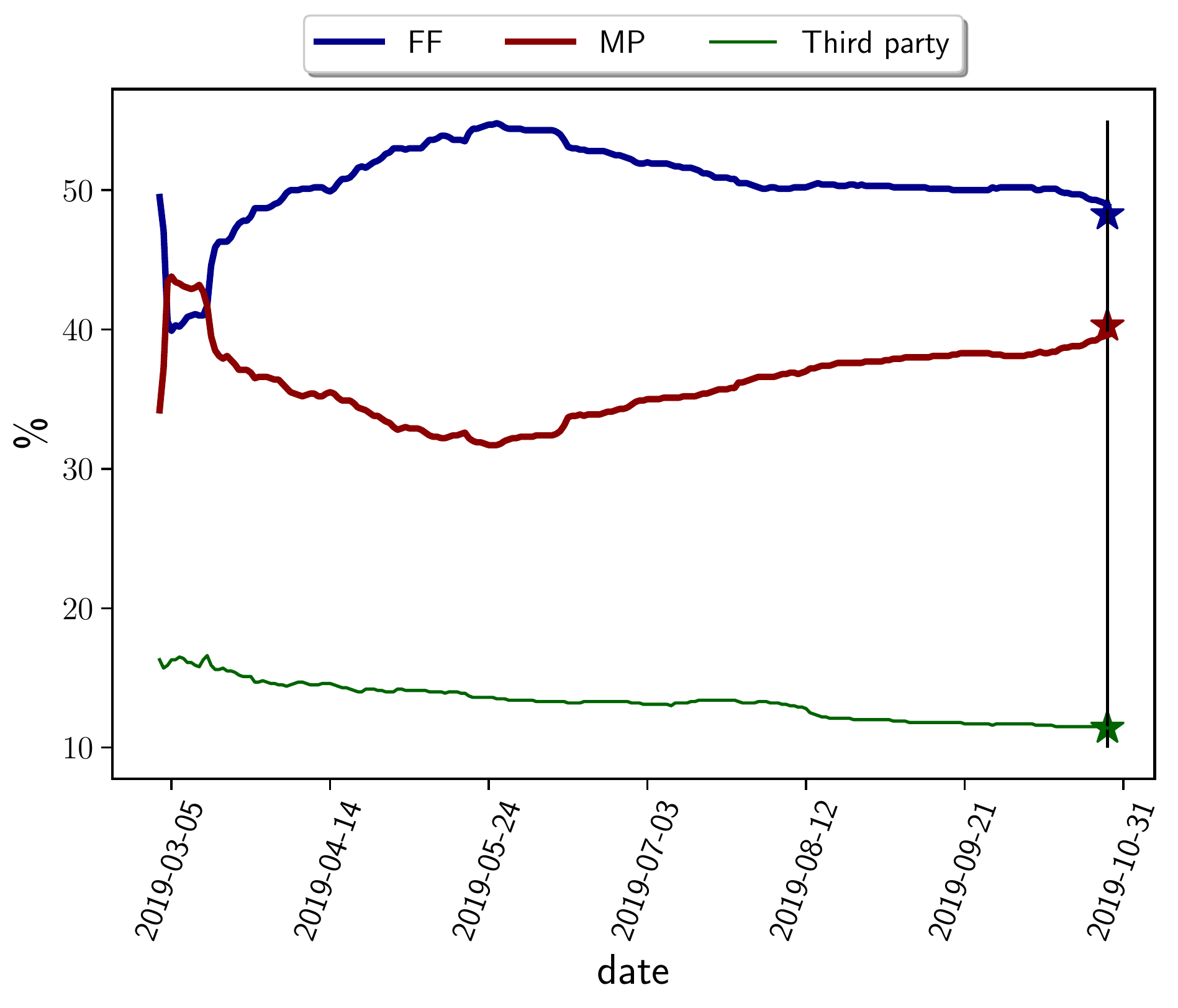}
  \caption{Cumulative MODEL 3 prediction of the AI model and
    comparison with primary on August 11, 2019 and general election
    results on October 27, 2019. Model 3 is the best fit to the real
    data.
  }
\label{fig:M3}
\end{figure}

\textsc{Model 3:} We use network homophily to determine the political
orientation of both undecided and unclassified users. FF(Undecided)
and FF(Unclassified) are considered supporters of Fern\'andez and
MP(Undecided) and MP(Unclassified) supporters of
Macri. Undecided(Undecided), Unclassified(Unclassified) belong to the
third party, see Fig~\ref{fig:M3}.

In the next section we compare the performances of these models on the
Argentina election.

\section{AI-BASED FORECAST FOR THE ARGENTINIAN ELECTION}

The models introduced so far allow us to define the daily supporters
of each candidate according to their retweet activity. Indeed
supporters are defined not simply according to the classification of
the majority of their retweet, but on the basis of the loyalty classes
they belong. Similarly to the simple tweets classification, we can
define for each model an instantaneous (window average) and a
cumulative (average) opinion.

Fig.~\ref{fig:14days} shows that an instantaneous indicator provides
an approximate fitting to the results of polls. We have already used
this indicator, in our previous study of the 2016 US presidential
election, to precisely fit the New York Times Aggregator of Polls at
The Upshot' \cite{bovet,nytimes}. This aggregator unifies a thousands
polls and weight them with proprietary information to produce a
weighted average of all the most trustable pollster in USA. While this
analysis is interesting and give the opportunity to predict
instantaneous changes in electoral opinion, this indicator does not
provide the electorate opinion as a whole and it is not the most
important predictor of the election outcome. It is not the greatest
information that can be extracted from social networks, either, and
indeed, it failed to predict the US 2016 election and the present
Argentina 2019. The estimator that predictor better the election is
provided when we consider the cumulative number of users from the
beginning of measurements, and not just the behavior of the users in a
small window of observation.

For this reason here we directly focus on the cumulative prediction
for the models introduced in the previous section. Table \ref{tab:mp}
reports the prediction of each model right before the day of the
general election day: October 27, 2019. The official results saw the
victory of Fern\'andez with 48.24\%. Macri scored 40.28\% and the
Third Party with 19.48\%. The average predictions obtained by
averaging the results of the five models are consistent (inside the
standard error) with the official outcome. Indeed, we obtain (49.3
$\pm$ 2.1)\% for FF, (36.8 $\pm$ 2.9)\% for MP and (13.9 $\pm$ 4.8)\%
for the Third Party. This in an outstanding result which highlights
the importance of considering loyalty classes for Political
elections. In order to establish the best among the five models we
compute the mean absolute error between each model' prediction and the
final results. Let $Y=\left\lbrace y_c \right\rbrace$ with $c \in
\left\lbrace FF,MP,TP \right\rbrace$ be the prediction of one model
and let $X=\left\lbrace x_c \right\rbrace$ with $c \in \left\lbrace
FF,MP,TP \right\rbrace$ be the official results. We define the MAE$_i$
(mean absolute error) for model $i$ as $\frac{\sum_{i \in c}
  (|x_i-y_i|)}{3}$.

\begin{figure}[!ht]
  \centering
  \subfloat[]{\includegraphics[width=.5\textwidth]{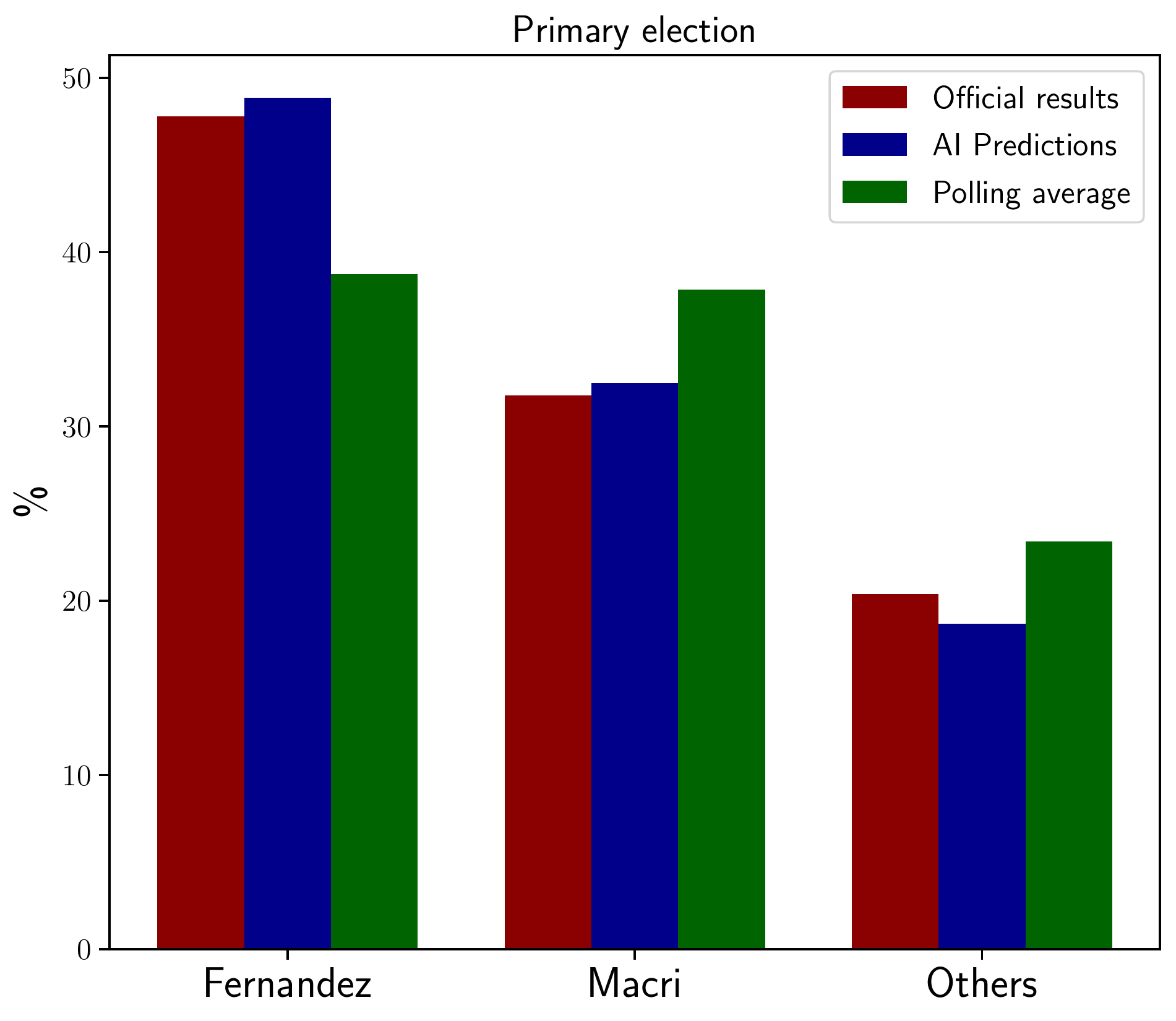}}
  \hfil
  \subfloat[]{\includegraphics[width=.5\textwidth]{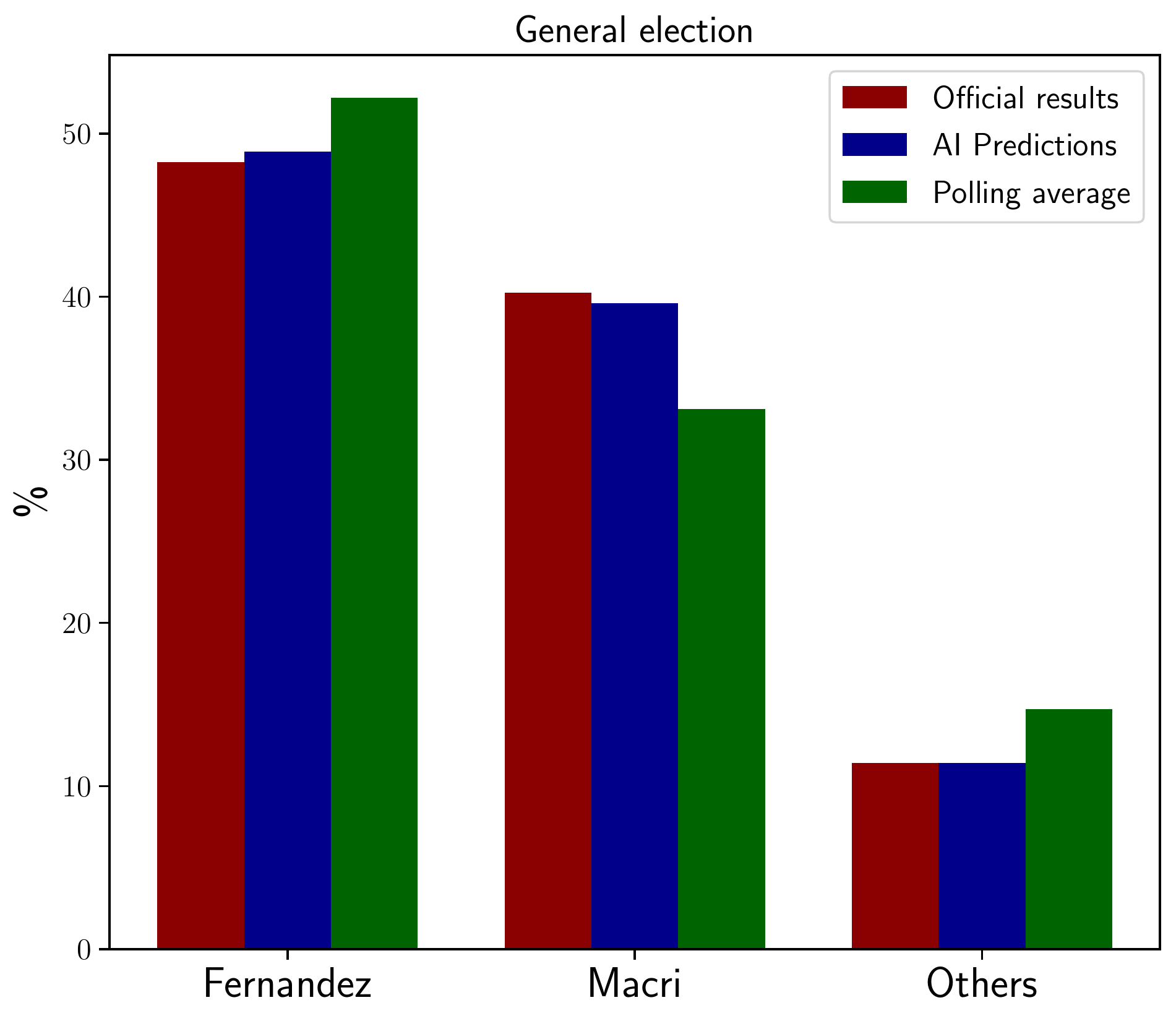}}
  \caption{Comparison between polling average (green), official results (red) and our prediction (blue) for both the (a) Primary (2019-08-11) and the (b) General election (2017-10-27).}
  \label{fig:comparison}
\end{figure}

Table~\ref{tab:mp} shows the MAE for each model. \textsc{Model 3},
based on the homophily detection for the undecided is the best
predictor with a mean absolute error of 0.53. This model predicted
48.9\% for FF (an overestimation of 0.66 points if compared to the
official result), 39.6\% for MP (an underestimation of 0.68 points)
and 11.6\% for the Third Party. As a matter of fact, the AI model
introduced so far is capable of predicting the Argentinian general
elections, by giving a percentage of electors for each candidate close
to the official one and outperforming traditional polls methods. See
Fig. \ref{fig:comparison}b.

Maybe the most important result is the performance of our algorithm
before the PASO, where all the pollsters failed too predict the +16\%
points difference between the two candidates (by strongly
underestimating their gap). \textsc{Model 3} predicts a difference of
almost 18\% points in favor of FF, close to the official result.

While the PASO results appears of secondary importance, they play a
central role in the Argentina political campaign and they are the most
difficult to guess because it's the first time the citizens officially
expressed their opinion on the election. Figure \ref{fig:comparison}
shows how traditional pollsters modified their prediction after the
primary elections, somehow fitting them with the PASO results. How
they modified their predictions is still not clear and some of the
pollsters (Elypsis for example) did not release any prediction after
the PASO.

A study of the hashtags and queries of the followers of the FF formula indicates that the vast majority of the people focused more on the poor economic situation in which the country was instead of the judicial cases of corruption that affect the FF candidates. Most of the hashtags reflect sentiment of hunger, chaos, crisis and despair. On the other hand, the expression of the followers of Macri-Pichetto is reflected in hashtags to give strength to the president but they do not reflect a feeling for the economic and political situation, but more a moral support, perhaps of resignation. The followers of Macri do not express too much their concerns about judicial cases of corruption either.

Finally, let us notice that the cumulative average depends on the initial time $T_0$. This value determines the initial fluctuations of the cumulative average,  which generally stabilize into a value that it is difficult to change unless a big swing in opinion of the electorate. To investigate this effect, we have recalculated the cumulative average by changing the origin of measurement $T_0$ in Fig. As we see from this figure, the predictions for the general elections cluster around the same value. We use this fluctuations to compute the error associate to our final predictions. We define the error as the standard deviation over the results of different realizations with $t<T_0$. Regarding \textsc{Model 3}, the estimated average error is 0.53\%. This result strengths the goodness of our prediction, consistent, inside the error bars, with the final results.

\section{Conclusion}

One of the fundamental tools of artificial intelligence in social
networks is that it captures changes in people's opinions without any
intervention and for an extended time. Then AI can capture the
sentiment of the millions of users who constantly express themselves
on the internet and change or maintain their positions. AI can also
filter this information from manipulators and bots and can reduce it
to its essence, by overcoming the problems traditional pollsters face:
low response rate, social desirability biases and the
mis-representation of the population.

The results of our analyses show that AI applied to big-data can be
used to successfully understand people' opinions over time. The
possibility of following the opinion of the same people through time,
and therefore the chance of defining loyalty classes is a fundamental
step in order to make good predictions. AI allows both to get the
percentage of supporters toward a candidate and reveals what is behind
these numbers, giving an idea of people sentiments. This is of
particular importance when one of the candidates is a controversial
politician and can generate different feelings leading to strong
polarization and biased responses to pollsters, which are not trusted
anymore by the great majority of people.

We expect that in the future traditional surveys may be incrementally
replaced by these new non-intrusive methods. AI is a thermometer that
provides the key to predicting not only the elections but the great
trends that develop at the local and global levels. We have shown how
AI allows to synthesize the opinion of millions of people including
those silent majorities of hidden voters who would not be heard
otherwise. We must not ignore that people are tired of answering
surveys. AI can then deduce, predict, interpret and understand what
people want to express.

\clearpage

\noindent

{\bf Acknowledgements:} GC acknowledges support from EU project,
HUMANE-AI-NET (grant number 952026). HAM owns shares of Kcore
Analytics.

\clearpage

\clearpage

\clearpage

\begin{table}[h!]
  \begin{center}
    \begin{tabular}{|l|ccc|cc|} 
    \hline%
      \textbf{Who will you vote?} &  \multicolumn{3}{c|}{\textbf{Who did you vote?}}  & \multicolumn{2}{c|}{\textbf{Was Pre PASO vote true?}}\\
       & AF-CFK & MM-MP & Other &  Yes & No \\
            \hline%
            \textbf{AF-CFK} & $91\%$ & $2\%$ & $8\%$ & $91\%$ & $9\%$ \\ 
            \hline%
             \textbf{MM-MP} & $6\%$ & $83\%$ & $11\%$ & $83\%$ & $17\%$ \\ 
            \hline%
             \textbf{Lavagna} & $19\%$ & $9\%$ & $72\%$ & $56\%$ & $44\%$ \\  
            \hline%
            \textbf{Del Cano} & $25\%$ & $0\%$ & $75\%$ & $54\%$ & $46\%$ \\ 
            \hline%
            \textbf{Espert} & $19\%$ & $14\%$ & $67\%$ & $53\%$ & $47\%$ \\ 
            \hline%
            \textbf{Gomez Centurion} & $10\%$ & $8\%$ & $83\%$ & $69\%$ & $31\%$ \\ 
            \hline%
            \textbf{Blank or Null} & $23\%$ & $4\%$ & $73\%$ & $47\%$ & $53\%$ \\ 
            \hline%
            \textbf{Unknown or Others} & $53\%$ & $11\%$ & $36\%$ & \* & \* \\  \hline
    \end{tabular}
    \caption{Vote disclosure analysis: ``Who are you going to vote in the PASO'' with ``Who did you vote in the PASO'' - using the same sampling and post-stratification methodology than in the Pre-PASO survey  \cite{Elypses}.}
\label{table-longitudinal}
  \end{center}
\end{table}

\begin{table}[h!]
  \begin{center}
    \begin{tabular}{|l|c|c|} 
    \hline%
      \textbf{} &  \textbf{Revealed}  & \textbf{Not Revealed}\\
            \hline%
            \textbf{Man} & $83\%$ & $17\%$  \\ 
            \hline%
             \textbf{Woman} & $81\%$ & $19\%$  \\ 
            \hline%
             \textbf{Between 16 and 30} & $67\%$ & $33\%$ \\  
            \hline%
            \textbf{Between 31 and 50} & $87\%$ & $13\%$  \\ 
            \hline%
            \textbf{Between 51 and 65} & $90\%$ & $10\%$  \\ 
            \hline%
            \textbf{More than 65} & $89\%$ & $11\%$  \\ 
            \hline%
            \textbf{Full Secondary} & $81\%$ & $19\%$  \\ 
            \hline%
            \textbf{Incomplete Secondary} & $81\%$ & $19\%$  \\ 
            \hline%
            \textbf{Full or incomplete Univ.} & $86\%$ & $14\%$  \\ 
            \hline%
            \textbf{Total} & $82\%$ & $18\%$  \\  \hline
    \end{tabular}
    \caption{Hidden vote by demographics \cite{Elypses}.}
    \label{table2}
  \end{center}
\end{table}

\begin{table}[h!]
  \begin{center}
    \begin{tabular}{|l|cccc|cccc|} 
    \hline%
            \textbf{Image $\%$ of the total}& \multicolumn{4}{c|}{\textbf{ Revealed}} & \multicolumn{4}{c|}{\textbf{Not Revealed}} \\
            \hline%
            \textbf{} & Positve  & Negative  & Regular &  NS/NC & Positve  & Negative  & Regular &  NS/NC \\ 
            \hline%
            \textbf{CFK} & $45\%$ & $43\%$ & $6\%$ & $5\%$ & $20\%$ & $48\%$& $22\%$ & $11\%$ \\ 
            \hline%
             \textbf{MM} & $36\%$ & $50\%$ & $10\%$ & $4\%$ & $26\%$ & $39\%$& $21\%$& $14\%$\\ 
            \hline%
             \textbf{AF} & $45\%$ & $39\%$ & $7\%$ & $8\%$ & $15\%$  &$28\%$& $28\%$&  $35\%$\\  \hline
    \end{tabular}
    \caption{CFK, AF, and MM Image as \% of the total \cite{Elypses}.}
    \label{table3}
  \end{center}
\end{table}

\clearpage

\clearpage

\clearpage

\clearpage

\clearpage

\begin{table}[!ht]
  \begin{tabular}{l|cccccc}
  \toprule
  Model & Precision (FF) & Recall (FF) & F1 (FF) & Precision (MP) & Recall (MP) & F1 (MP) 	  \\ \colrule
  LR & 0.83      & 0.83   & 0.83  & 0.83   & 0.83   & 0.83   \\ \colrule
  SVM & 0.81     & 0.81   & 0.81  & 0.81   & 0.80   & 0.81   \\ \colrule
  NB & 0.79      & 0.80   & 0.80  & 0.80   & 0.79   & 0.80   \\ \colrule
  RF & 0.74      & 0.80   & 0.77  & 0.79   & 0.72   & 0.75   \\ \colrule
  DT & 0.76      & 0.76   & 0.76  & 0.76   & 0.76   & 0.76   \\ \botrule
  \end{tabular}
  \caption{Performance of the classification models: Logistic Regression (LR),Supporting Vector Machine (SVM), Naive Bayes (NB), Random Forest (RF) an Decision Tree (DT).}
  \label{tab:accuracy}
\end{table}
\clearpage

\clearpage

\clearpage

\clearpage

\begin{table}[!ht]
\centering
\begin{adjustbox}{max width=\textwidth}
\begin{tabular}{c|l}
\toprule
MODEL 1 & Supporters (Users) \\ [0.5ex] 
\colrule
FF          & Ultra loyal MP, loyal MP$\to$MP, loyal MP$\to$FF and loyal MP$\to$Undecided   \\ \colrule
MP          & Ultra loyal FF, loyal FF$\to$FF, loyal FF$\to$MP, loyal FF $\to$ Undecided   \\ \colrule
Third party & Undecided$\to$MP, Undecided$\to$FF, Undecided$\to$Undecided, Unclassified   \\ \colrule
MODEL 2 &  \\ [0.5ex] 
\colrule
FF          & Ultra loyal FF, loyal FF$\to$FF + loyal FF$\to$MP + loyal FF$\to$Undecided, FF(undecided)  \\ \colrule
MP          &  Ultra loyal FF, loyal FF$\to$FF, loyal FF$\to$MP, loyal FF $\to$ Undecided, MP(undecided)  \\ \colrule
Third party &  Undecided(undecided), Unclassified   \\ \colrule
MODEL 3 &  \\ [0.5ex] 
\colrule
\botrule
\end{tabular}
\end{adjustbox}
\caption{The definition of three models according opinion modeling.}
\label{tab:models}
\end{table}

\begin{table}[!ht]
\setlength{\tabcolsep}{6mm}{
\begin{tabular}{c|cccc}
\toprule
MODEL & FF (\%)   & MP (\%)   & Third party (\%)   & MAE (\%) \\ \colrule
1     & 45.9 & 32.5 & 21.6 & 6.71\\ \colrule
2     & 49.1 & 34.0 & 16.8 & 4.21\\ \colrule
3     & 48.9 & 39.6 & 11.5 & 0.53\\ \colrule

\end{tabular}}
\caption{Models' prediction results for the general election on October 27.}
\label{tab:mp}
\end{table}



\clearpage

\centerline{Supplementary Information}

\begin{table}[!ht]
\setlength{\tabcolsep}{7mm}{
\begin{tabular}{l|cc}
\toprule
Tweets              & Bots   & Users    \\ \colrule
Total               & 538359 & 67336507 \\ \colrule
Daily				& 2243   & 280568 \\ \colrule
Daily classified    & 1617   & 211229   \\ \colrule
Daily classified MP & 619    & 114653   \\ \colrule
Daily classified FF & 998    & 96576    \\ \botrule
\end{tabular}}
\caption{Tweets statistics. We report the Total number of tweets collected.
 The average daily number of tweets classified (Daily classified) and the average daily classified tweets for each candidate.}
\label{tab:tweets_stat}
\end{table}

\begin{table}[!ht]
\setlength{\tabcolsep}{7mm}{
\begin{tabular}{l|cc}
\toprule
Users            & Bots   & Users      \\ \colrule
Total               & 17953 & 2252551  \\ \colrule
Daily				& 732   & 83330 \\ \colrule
Daily classified    & 560   & 63808    \\ \colrule
Daily classified MP & 198    & 31497   \\ \colrule
Daily classified FF & 362    & 32310   \\ \botrule
\end{tabular}}
\caption{Users statistics. We report the Total number of users collected.
 The average daily number of users classified (Daily classified) and the average daily classified users for each candidate.}
\label{tab:users_stat}
\end{table}



\begin{figure}[!ht]
  \includegraphics[width=.8\textwidth]{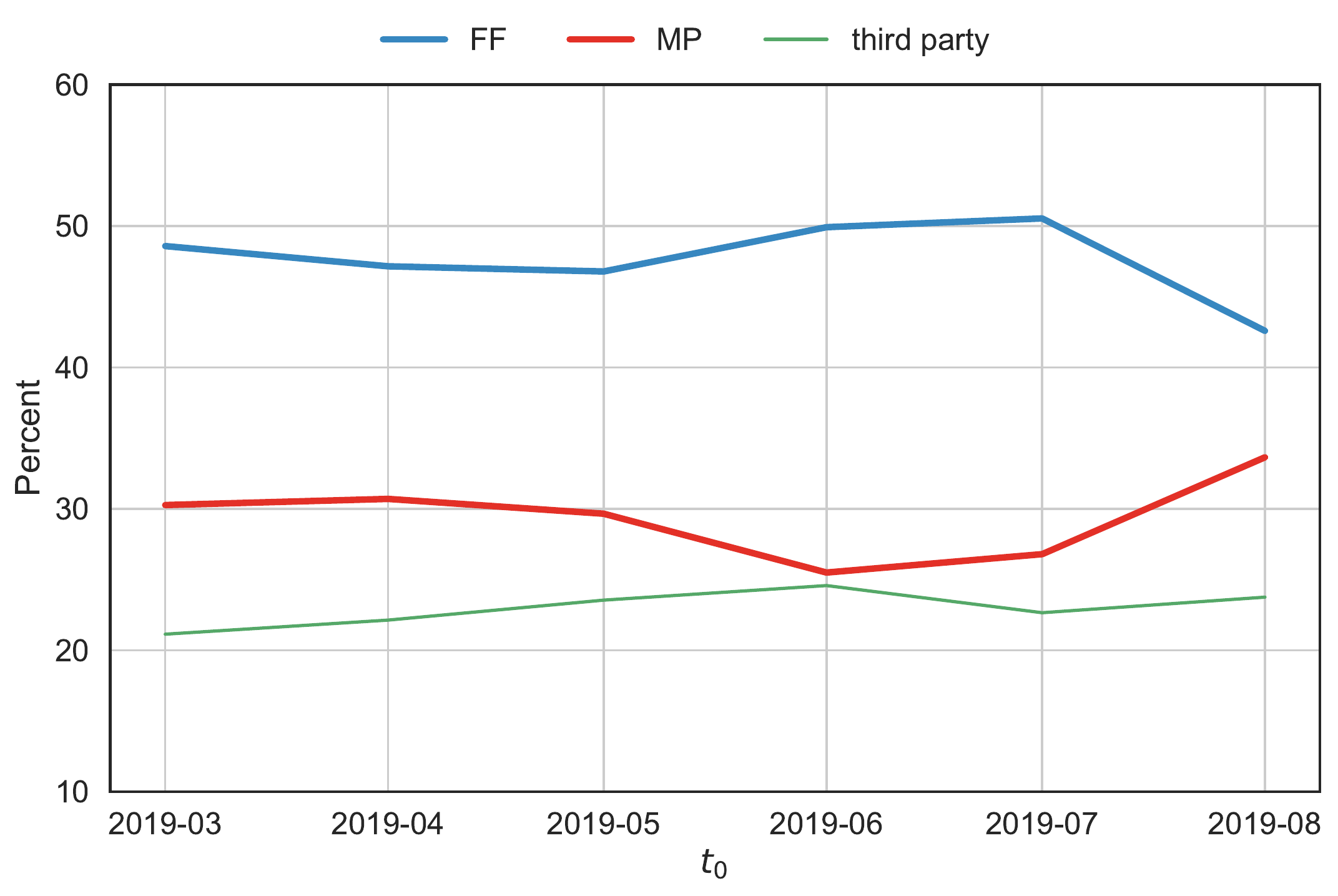}
  \caption{Cumulative prediction with $T_0=$March, 1 until few days (August 1) before PASO.}
  \label{fig:cumulative}
\end{figure}

\clearpage
\begin{longtable}
{ p{.10\textwidth} p{.40\textwidth} p{.10\textwidth} p{.10\textwidth}} 
\toprule
\#  & Hashtag                       & Camp & Count  \\ \colrule
1 & sisepuede & M & 216757 \\ \colrule
2 & macri & M & 102172 \\ \colrule
3 & albertopresidente & K & 65644 \\ \colrule
4 & axelgobernador & K & 60920 \\ \colrule
5 & juntosporelcambio & M & 52617 \\ \colrule
6 & yovotomm & M & 49557 \\ \colrule
7 & ladamosvuelta & M & 47546 \\ \colrule
8 & cfk & K & 38718 \\ \colrule
9 & cambiemos & M & 38538 \\ \colrule
10 & sevan & K & 32110 \\ \colrule
11 & habraconsecuencias & K & 28406 \\ \colrule
12 & 24a & M & 23819 \\ \colrule
13 & macrihacetecargo & K & 21364 \\ \colrule
14 & novuelvenmas & M & 21054 \\ \colrule
15 & cronicaanunciada & K & 20293 \\ \colrule
16 & frentedetodos & K & 18115 \\ \colrule
17 & albertoycristina & K & 17482 \\ \colrule
18 & sinceramente & K & 15403 \\ \colrule
19 & juntossomosimparables & M & 14405 \\ \colrule
20 & sevanenprimeravuelta & K & 14091 \\ \colrule
\caption{The top 25 hashtags from March and July in 2019. The camp field represents the classification: M stays for Macri and K for Fern\'andez (from the name of the running
mate Cristina Fern\'andez de Kirchner. Count indicates the number of time a given hashtag appears in the dataset.}
\label{hashtag_table_March}
\end{longtable}

\clearpage
\begin{longtable}
{ p{.10\textwidth} p{.40\textwidth} p{.10\textwidth} p{.10\textwidth}} 
\toprule
\#  & Hashtag                         & Camp & Count  \\ \colrule
1 & cambiemos & M & 150754 \\ \colrule
2 & macri & M & 120363 \\ \colrule
3 & cfk & K & 63052 \\ \colrule
4 & sinceramente & K & 56198 \\ \colrule
5 & habraconsecuencias & K & 52896 \\ \colrule
6 & novuelvenmas & M & 39352 \\ \colrule
7 & juntosporelcambio & M & 35186 \\ \colrule
8 & macritevuelvoaelegir & M & 32241 \\ \colrule
9 & cronicaanunciada & K & 30954 \\ \colrule
10 & massa & K & 29565 \\ \colrule
11 & axelgobernador & K & 29039 \\ \colrule
12 & navarro2019 & K & 27885 \\ \colrule
13 & defensoresdelcambio & M & 27010 \\ \colrule
14 & fuerzacristina & K & 21342 \\ \colrule
15 & andatemacri & K & 19486 \\ \colrule
16 & chaumacri & K & 19184 \\ \colrule
17 & debodecir & M & 19173 \\ \colrule
18 & hayotrocamino & K & 19152 \\ \colrule
19 & mm2019 & M & 16117 \\ \colrule
20 & sracristinalecuentoque & M & 16082 \\ \colrule
\caption{The top 25 hashtags from August and October in 2019. The camp field represents the classification: M stays for Macri and K for Fern\'andez (from the name of the running mate Cristina Fern\'andez de Kirchner. Count indicates the number of time a given hashtag appears in the dataset.}
\label{hashtag_table_August}
\end{longtable}

\end{document}